\documentclass{sig-alternate}
\usepackage{mathptmx} 

\usepackage{fancyhdr}
\usepackage[normalem]{ulem}
\usepackage[hyphens]{url}
\usepackage[sort,nocompress]{cite}
\usepackage[final]{microtype}
\usepackage{flushend}
\usepackage{balance}

\usepackage[bookmarks=true,breaklinks=true,colorlinks,linkcolor=black,citecolor=blue,urlcolor=black]{hyperref}

\usepackage[noend]{algorithmic}
\usepackage{algorithm}
\usepackage{xspace}
\usepackage{listings}
\usepackage{tikz}
\usepackage{makecell}

\usepackage{tabularx}
\usepackage{booktabs}
\usepackage{float}

\def\e#1{\emph{#1}}

\newcommand{\mparagraph}[1]{\vskip 0.4em\noindent\textbf{#1.}\,\,}
\newcommand{\hide}[1]{}

\newcommand{\fig}[1]{Figure~\ref{#1}}
\newcommand{\tab}[1]{Table~\ref{#1}}
\newcommand{\sect}[1]{Section~\ref{#1}}
\newcommand{\step}[1]{\circleHig{#1}}

\definecolor{midnightblue}{rgb}{0.1, 0.1, 0.44}
\definecolor{darkolivegreen}{rgb}{0.33, 0.42, 0.18}

\DeclareRobustCommand\circleCtj[1]{\tikz[baseline=(char.base)]{\node[shape=circle,draw,inner sep=0.5pt] (char) {\small #1};}}

\DeclareRobustCommand\circleCup[1]{\tikz[baseline=(char.base)]{\node[white, shape=circle,draw,inner sep=0.5pt,fill=black] (char) {\small #1};}}
\DeclareRobustCommand\circleMak[1]{\tikz[baseline=(char.base)]{\node[white, shape=circle,draw,inner sep=0.5pt,fill=darkolivegreen] (char) {\small #1};}}
\DeclareRobustCommand\circleLub[1]{\tikz[baseline=(char.base)]{\node[white,shape=circle,draw,inner sep=0.5pt,fill=blue] (char) {#1};}}
\DeclareRobustCommand\circleMid[1]{\tikz[baseline=(char.base)]{\node[white, shape=circle,draw,inner sep=0.5pt,fill=brown] (char) {\small #1};}}
\DeclareRobustCommand\circleHig[1]{\tikz[baseline=(char.base)]{\node[white, shape=circle,draw,inner sep=0.5pt,fill=midnightblue] (char) {\small #1};}}

\def\partitle#1{\mparagraph{#1}}

\newcolumntype{L}[1]{>{\raggedright\arraybackslash}m{#1}}
\newcolumntype{C}[1]{>{\centering\arraybackslash}m{#1}}
\newcolumntype{R}[1]{>{\raggedleft\arraybackslash}m{#1}}

\def\sysname {TrieJax\xspace}
\def \lub {\e{LUB}\xspace}
\def \mmaker {\e{MatchMaker}\xspace}
\def \midwife {\e{Midwife}\xspace}
\def \cupid {\e{Cupid}\xspace}
\def \irs {{partial join results cache}\xspace}
\def \tirs {{PJR cache}\xspace}

\newcommand{\algname}[1]{{\sf #1}}
\def\myrulewidth{3.20in}

\def\therule{\rule{\myrulewidth}{0.2pt}}

\newenvironment{algseries}[2]
{\centering\begin{figure}[#1]\begin{center}\def\thecaption{\caption{#2}}
\vskip-0.8em\begin{tabular}{p{\myrulewidth}}\therule\end{tabular}\vskip0.2em}
{\thecaption \end{center}\end{figure}}

\newenvironment{insidealg}[2]
{\normalsize\begin{insidecode}{#1}{#2}{Algorithm}}
{\end{insidecode}}

\newenvironment{insidecode}[3]
{
\begin{tabular}{p{\myrulewidth}}
\multicolumn{1}{c}{\rule{0mm}{3mm}{\bf #3} $\algname{#1}(\mbox{#2})$\vspace{-0.6em}}\\
\therule\vskip-0.8em\therule
\vspace{-1em}
\begin{algorithmic}[1]}
{\end{algorithmic}
\vskip-0.3em\therule
\end{tabular}}

\def\cache{\mathit{cache}}

\fancypagestyle{firstpage}{
  \fancyhf{}
  
  \fancyfoot[C]{\thepage}
}

\title{The \sysname Architecture: Accelerating Graph Operations Through Relational Joins}

\begin{document}

\numberofauthors{3}

\author{
%
%
\alignauthor
Oren Kalinsky\\
       \affaddr{Technion, Israel}
\alignauthor
Benny Kimelfeld\\
       \affaddr{Technion, Israel}
\alignauthor
Yoav Etsion\\
       \affaddr{Technion, Israel}
}




\maketitle
\thispagestyle{firstpage}
\pagestyle{plain}


\begin{abstract}
	
	Graph pattern matching (e.g., finding all cycles and cliques) has 
	become an important component in many critical domains such as 
	social networks, biology and cyber-security.
	This development motivated research to develop faster algorithms 
	that target graph pattern matching.
	In recent years, the database community has shown that mapping 
	graph pattern matching problems to a new class of relational join 
	algorithms provides an efficient framework for computing these 
	problems.
	
	In this paper, we argue that this new class of relational join 
	algorithms is highly amenable to specialized hardware 
	acceleration thanks to two fundamental properties: improved 
	locality and inherent concurrency. 
	The improved locality is a result of the provably bound number of 
	intermediate results these algorithms generate, which results in 
	smaller working sets. 
	In addition, their inherent concurrency can be leveraged for 
	effective hardware acceleration and hiding memory latency.
	
	We demonstrate the hardware amenability of this new class of 
	algorithms by introducing \sysname, a hardware accelerator for 
	graph pattern matching. 
	The \sysname design leverages the improved locality and high 
	concurrency properties to dramatically accelerate graph pattern 
	matching, and can be tightly integrated into existing manycore 
	processors. 
	We evaluate \sysname on a set standard graph pattern matching 
	queries and datasets. 
	Our evaluation shows that \sysname outperforms recently proposed 
	hardware accelerators for graph and database processing that do 
	not employ the new class of algorithms by $7-63 \times$ on 
	average (up to $539\times$), while consuming $15-179 \times$ less 
	energy (up to $1750 \times$). 
	systems that do incorporate modern relational join algorithms by 
	$9-20 \times$ on average (up to $45 \times$), while consuming 
	$59-110 \times$ less energy (up to $372 \times$).
\end{abstract}


%

\section{Introduction}
Analyzing the relationships in a graph has become a key building block in many domains, including social networks~\cite{Wilson:2009:UIS:1519065.1519089}, biology~\cite{DBLP:conf/ismb/IdekerOSS02}, and artificial intelligence~\cite{shadbolt2006semantic}. 
A recent study~\cite{Sahu:2017:ULG:3186728.3164139} analyzed the common challenges in graph processing and found that pattern matching problems, namely finding all instances of a given pattern in a graph, to be a dominantly popular problem in graph application domains. 
Graph pattern matching problems, however, are known to be computationally intensive and thus challenge algorithm designers.


In recent years, the database community has proposed new relational join algorithms that efficiently map graph problems to query evaluation over relational databases. In particular, 
a new breed of \emph{Worst-Case Optimal Join} (WCOJ) algorithms
has been introduced and studied~\cite{DBLP:conf/pods/NgoPRR12, DBLP:conf/icdt/Veldhuizen14, AboKhamis:2016:FQA:2902251.2902280, DBLP:conf/pods/NgoNRR14, DBLP:conf/edbt/KalinskyEK17}.
These algorithms were shown to be theoretically superior to the traditional join algorithms~\cite{DBLP:journals/siamcomp/AtseriasGM13}, and WCOJ-based solutions for graph matching problems have been shown to deliver superior performance compared to known graph algorithms~\cite{DBLP:conf/sigmod/NguyenABKNRR14}. 

From an architectural perspective,
pattern matching via WCOJ algorithms offers two features that make it hardware friendly.
First, WCOJ algorithms bound the number of intermediate results they generate, which greatly reduces the algorithms' working-set size, reduces data transfers to and from memory, and makes them more amenable to caching.
In contrast, traditional join algorithms partition multi-way joins (a join operation on multiple relations, or tables) into a tree of binary join operations. Each binary join operation then performs a memory scan on its input relations and generates a (potentially huge) intermediate relation. Importantly, many of the intermediate results are typically filtered out by subsequent join operations.
Second, WCOJ algorithms are highly concurrent. Although the algorithms'  control flow is non-trivial, which makes them unsuitable for GPGPUs, their inherent concurrency makes them amenable to hardware acceleration. In addition, this inherent concurrency allows specialized accelerators to apply multithreading techniques to hide memory latency.

In this paper, we argue that WCOJ-based algorithms for solving graph pattern matching problems are highly amenable to specialized acceleration.
We describe how graph problems can be mapped to relational join operations and detail how the new breed of relational join algorithms maps to hardware.
We further present \sysname, an on-die, domain-specific accelerator that leverages the hardware-friendly properties of WCOJ algorithms to accelerate graph pattern matching problems and dramatically reduce energy consumption.

\sysname employs a highly-concurrent WCOJ variant~\cite{DBLP:conf/edbt/KalinskyEK17} that scans table indexes stored in a tree-based data-structures. 
In addition, \sysname caches partial join results to speed up computation and minimize memory traffic.
Furthermore, \sysname is designed to be small enough to serve as a dedicated core in a standard many-core processor. This design allows \sysname to use the same memory system as other cores in the system, a property that is crucial for operating on large data sets. This is in contrast to discrete accelerators (e.g., GPGPUs), which have limited directly-attached memory capacity and require frequent data transfers between the system memory and their local memory.
We validated the performance, area, and energy consumption of \sysname using a synthesized and placed\&routed RTL implementation, whose results drove a cycle-accurate simulator.

We demonstrate the improved performance and reduced energy obtained by leveraging WCOJ-based graph pattern matching algorithms 
via a comparison of  \sysname to recently proposed database and graph processing accelerators~\cite{Wu:2014:QAD:2654822.2541961, Ham:2016:GHE:3195638.3195707}.
Unlike \sysname, these accelerators employ traditional join algorithms and are therefore susceptible to the potential explosion of intermediate results that these algorithms generate.
We also demonstrate the benefits of hardware acceleration of WCOJ algorithms by comparing \sysname to two software database management systems~\cite{Aberger:2016:ERE:2882903.2915213,DBLP:conf/edbt/KalinskyEK17} that are based on the new breed of WCOJ algorithms.
Ultimately, \sysname outperforms the state-of-the-art hardware accelerators by $7-63 \times$ on average, while consuming $15-179 \times$ less energy. Compared to the software systems, \sysname runs $9-20 \times$ faster and consumes $59-110 \times$ less energy.

In summary, we make the following contributions:
\begin{itemize}
    \item We present the hardware-friendly properties of WCOJ algorithms and how they can be used for solving graph matching problems.
    \item We present a domain-specific hardware accelerator that leverages the hardware-friendly properties of WCOJ algorithms for graph pattern matching.
    \item We demonstrate the performance and energy benefits of a WCOJ-based pattern matching accelerator through a detailed experimental evaluation.
\end{itemize}

The rest of the paper is organized as follows. 
\sect{sec:background} explains how graph pattern matching algorithms are mapped to relational join operations and describe the new breed of WCOJ algorithms in more detail. 
Section~\ref{sec:arc} describes the \sysname accelerator architecture and how it maps a WCOJ algorithm to hardware.
Finally, Section~\ref{sec:eval} presents an experimental evaluation comparing the performance and energy consumption of \sysname to database and graph analytics accelerators on standard graph pattern matching queries. 


\section{New relational join algorithms and graph pattern matching} \label{sec:background}
Graph pattern matching problems have become paramount in different computational domains, including social networks~\cite{Wilson:2009:UIS:1519065.1519089}, biology~\cite{DBLP:conf/ismb/IdekerOSS02}, and artificial intelligence~\cite{shadbolt2006semantic}. 
Yet despite the focus in recent years on developing general graph analytics frameworks (including hardware accelerators and optimizations~\cite{Ham:2016:GHE:3195638.3195707, DBLP:conf/micro/MukkaraBAM018, DBLP:conf/fccm/NurvitadhiWWHNH14}), these frameworks mostly benefit more well-known graph problems (e.g., Breadth-First Search and PageRank) but often ignore graph pattern matching problems. 

Recent advances in relational database theory have introduced methods for efficient computation of graph algorithms based on relational join operations.
These methods have been found particularly effective for computing graph pattern matching problems.
In this section, we describe how relational join operations can be used for graph analytics and review the algorithmic advances in relational join algorithms.
We then discuss how these algorithms, unlike traditional ones, avoid generating a large number of intermediate results, which makes them amenable to hardware acceleration.

\hide{
\lstset{mathescape,columns=fullflexible,basicstyle=\ttfamily\small,frame=single}
\begin{figure}
    \centering
    \begin{minipage}[t]{0.95\columnwidth}
        \begin{lstlisting}
SELECT S.name, C.name, R.grade 
FROM Student as S, Registration as R, 
Course as C WHERE 
S.id = C.studentID and R.courseID = C.id
        \end{lstlisting}
    \end{minipage}
    \caption{A relational join query example \label{fig:rel_query}}
\end{figure}
}

{
\lstset{mathescape,columns=fullflexible,basicstyle=\ttfamily\small,frame=single}
\begin{figure}
    \centering
    \begin{minipage}[t]{0.98\columnwidth}
        \begin{lstlisting}
SELECT *
FROM Posts as R, Likes as S, Follows as T
WHERE R.postID=S.post and S.user=T.followed
        \end{lstlisting}
    \end{minipage}
    \caption{A relational join query example \label{fig:rel_query}}
    \vspace{-2ex}
\end{figure}
}

\subsection{Relational Join}
Relational database management systems (RDBMS) are a common solution for data management. This type of database follows the relational model of data. In this model, the data is stored in relations, also known as \e{tables} (e.g., 
the tables \e{Posts} and \e{Likes} mentioned in Figure~\ref{fig:rel_query}), the table columns are the attributes (e.g., \e{user} and \e{postID}), and the rows (tuples) are the values. Each row in the table can have an attribute that is a unique primary key. For instance, the \e{Posts} table has the primary key \e{postID}. Other tables can reference a specific table through its primary key using an attribute known as a \e{foreign key}. In our example, the \e{Likes} table can reference a post using the \e{post} attribute as a foreign key to \e{Posts}.

A relational join query is a query that analyzes the relationship between tables via shared attribute values. Figure~\ref{fig:rel_query} shows a simple relational join query---the natural join of the three relations \e{Posts}, \e{Likes} and \e{Follows}.  The query computes information about posts liked by users with followers. More precisely, the query is asking for tuples where each consists of a post, a user who likes the post, and a follower of the user.  While there are different types of join operations, in this work we focus on natural equi-joins where tables are joined by equality of mentioned attributes.

\partitle{Relational join for graph analytics}
Many graph algorithms can be translated to (SQL-like) join queries, which allows solving graph problems using RDMS solutions. 
A finite graph is commonly represented in an RDBMS by an adjacency list relation. Each row in the relation represents an edge between two vertices in the graph. Graph patterns queries are translated to join queries. For example, given a graph relation $G$, the query $Q(x,y,z) = G(x,y)\Join G(y,z) \Join G(z, x)$ returns all the triangles in the graph.

Mapping graph pattern matching algorithms to WCOJ-based systems have been shown to be highly effective, and WCOJ-based systems can provide speedups of up to two orders of magnitude compared to low-level graph analytic solutions~\cite{DBLP:journals/siamcomp/AtseriasGM13}. 
Nevertheless, these performance benefits are not universal across all graph algorithms. For example, Aberger et al.~\cite{Aberger:2016:ERE:2882903.2915213} have shown that problems such as SSSP~\cite{Johnson:1977:EAS:321992.321993} and PageRank~\cite{page1999pagerank}) do not enjoy similar benefits. 
In this paper we focus on the important family of graph pattern matching problems.

\partitle{Database and graph analytic acceleration}
Acceleration of database and graph analytics using SIMD, GPGPU, or specialized hardware accelerator has been a common interest in both the database and micro-architecture communities. Q100 by Wu et al.~\cite{Wu:2014:QAD:2654822.2541961} was the first hardware accelerator that fully supported relational operations. It incorporates relational operators (such as Sort, Select, and Merge-Join) as hardware components in a hardware accelerator for relational column stores. Q100 offers a solution that searches the best custom chip for a specific query from time and energy perspectives. Q100 achieves a speedup of $10 \times$ on TPC-H compared to MonetDB, a commonly used column store. Graphicionado by Ham et al.~\cite{Ham:2016:GHE:3195638.3195707} focuses on graph analytics. It implements the vertex-programming model in hardware with embedded programmable units that allow flexible support for graph algorithms such as PageRank or SSSP. The Graphicionado hardware accelerator achieves a speedup of $1.76 - 6.5 \times$ compared to GraphMat~\cite{Sundaram:2015:GHP:2809974.2809983}, a vertex-programming framework that scales using sparse matrix representation and MPI.

\partitle{New relational join algorithms for graph analytics}
Recently, the database community established new theoretical and algorithmic advances in the area of relational join algorithms. 
Given a join query with more than two relations, standard join algorithms use the pairwise join approach. The traditional algorithms, such as hash-join~\cite{dewitt1985multiprocessor} or sort-merge join~\cite{mishra1992join}, join two relations at a time to create a new intermediate relation. This intermediate relation will later be joined with another (input or intermediate) relation until the final result is computed. Recent work by Atserias et al.~\cite{DBLP:journals/siamcomp/AtseriasGM13} shows that pairwise join algorithms can generate many unnecessary intermediate results that are not part of the final result. It helps define a tight bound, called the AGM bound, on the maximum number of results that can be returned from a query in the worst case. 

We illustrate the AGM bound with an example. Given the relations $A$, $B$ and $C$, consider the triangle join query: $Q(x,y,z) = A(x,y)\Join B(y,z) \Join C(z, x)$. For simplicity, we assume that each relation has $N$ values. The AGM bound proves that the query result $Q(x,y,z)$ contains no more than $N^\frac{3}{2}$ results. However, pairwise join algorithms can generate an intermediate result with up to $N^2$ values, while many of them will be filtered by the third relation. Any join algorithm that provides the same complexity as the AGM bound is called Worst-Case Optimal Join (WCOJ) algorithm. In contrast, the traditional pairwise join algorithms are not worst-case optimal.
More formal definitions and extensions to general queries can be found in a full survey by Ngo et al.~\cite{Ngo:2014:SSB:2590989.2590991}.

\partitle{Systems and acceleration of WCOJ algorithms\label{subsec:related}}
Wu et al.~\cite{DBLP:conf/vldb/2014adms} used GPGPU to accelerate a WCOJ algorithm and achieved $2-6 \times$ speedups compared to CPU. 
EmptyHeaded by Aberger et al.~\cite{Aberger:2016:ERE:2882903.2915213} offered a relational query engine that maps Generic Join, a WCOJ algorithm, to parallel SIMD operations on a standard CPU. For graph algorithms, EmptyHeaded achieves comparable results to Galois~\cite{Nguyen:2013:LIG:2517349.2522739}, a low-level hand-tuned query engine comparable to GraphMat. On graph pattern matching queries, it achieves a $2-60 \times$ speedup compared to other CPU based solutions. 
Section~\ref{sec:eval} compares our \sysname to the systems above or their baselines.

\subsection{Cached TrieJoin \label{subsec:ctj}}
Following the publication of the AGM bound, a plethora of Worse Case Optimal Join (WCOJ) algorithms was published~\cite{DBLP:conf/pods/NgoPRR12, DBLP:conf/icdt/Veldhuizen14, AboKhamis:2016:FQA:2902251.2902280, DBLP:conf/pods/NgoNRR14, DBLP:conf/edbt/KalinskyEK17}. Experimental analysis by Nguyen et al.~\cite{DBLP:conf/sigmod/NguyenABKNRR14} has shown that the WCOJ algorithms provide significant speedups on complex join queries compared to the traditional approaches such as state of the art RDBMS, graph engines and column stores. 

LeapFrog TrieJoin~\cite{DBLP:conf/icdt/Veldhuizen14}, also known as LFTJ, is a commonly used WCOJ algorithm. Its main idea is to iterate over trie-based indexes of the relations in a backtracking manner to generate the join query results. LFTJ does not generate any intermediate results and thus yields a low memory footprint, but it does so at the expense of recomputing recurring intermediate partial joins. Furthermore, the recurring computations have little memory locality as they repeatedly scan irregular, tree-based tries.
Cached TrieJoin (CTJ) by Kalinsky et al.~\cite{DBLP:conf/edbt/KalinskyEK17} eliminates much of the recurring partial join computations by selectively caching partial join results using the available system memory (without violating the WCOJ property). This behavior can benefit low-memory environments and is therefore used as one of the main building blocks in our system.

CTJ operates as follows: Given a query, CTJ decomposes the query structure to detect which attributes can be valid keys and their respective cached values. Then, it uses the caching system to drive the TrieJoin, while saving partial join results in the cache and extracting them later to avoid recurrent computations. CTJ shows a $10 \times$ speedup on average compared to LFTJ, and even better speedups compared to traditional query engines.

\begin{figure}
    \includegraphics[width=\columnwidth,trim={53ex 30ex 25ex 30ex}, clip=true]{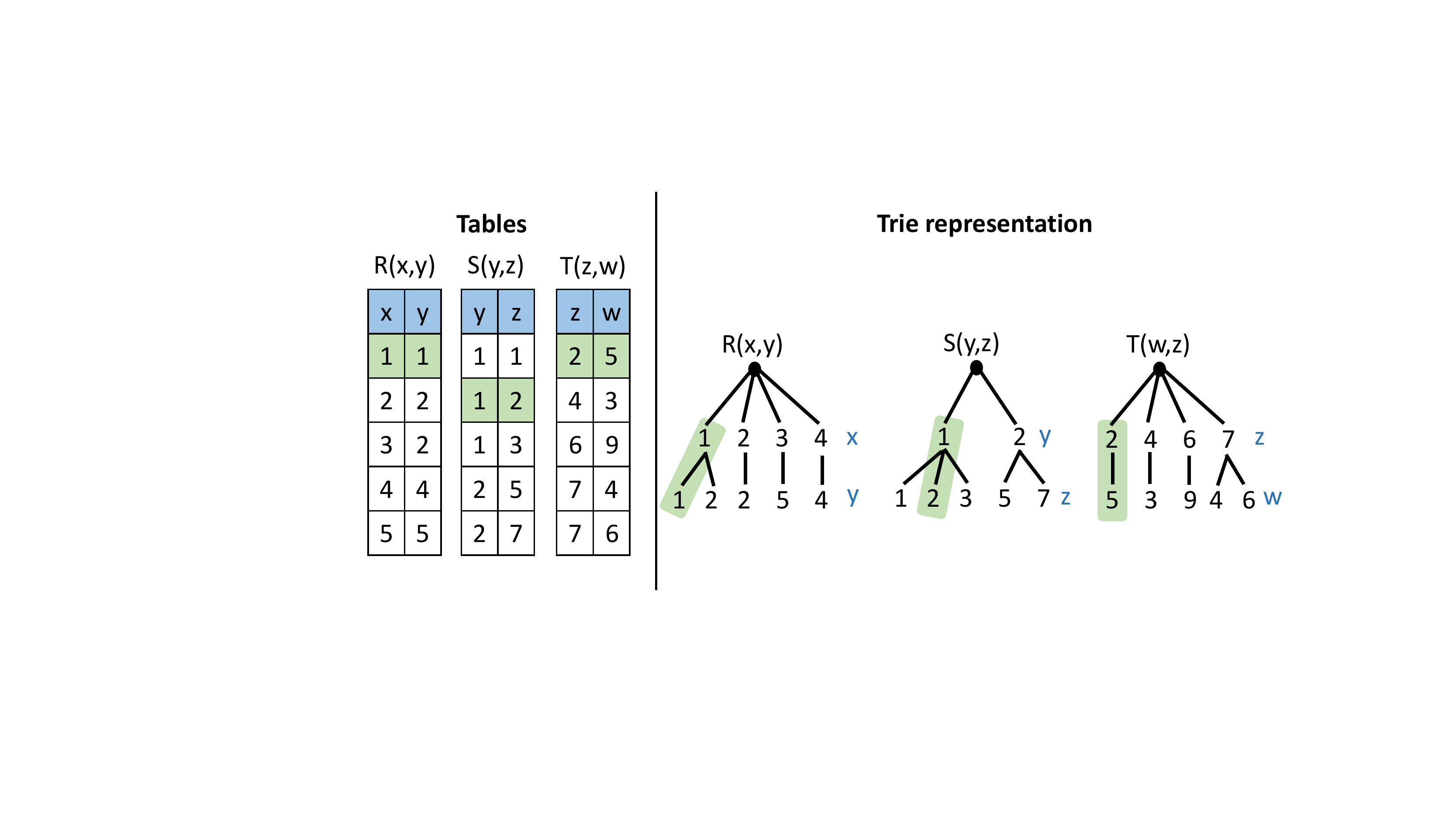}
    \caption{Example of tables from a social network (left) and their trie representation (right). Marked (green) a Path-4 join result between the three tables. \label{fig:index_ex}}
\end{figure}


\begin{figure*}
    \includegraphics[width=\textwidth,trim={0ex 5ex 0ex 10ex}, clip=true]{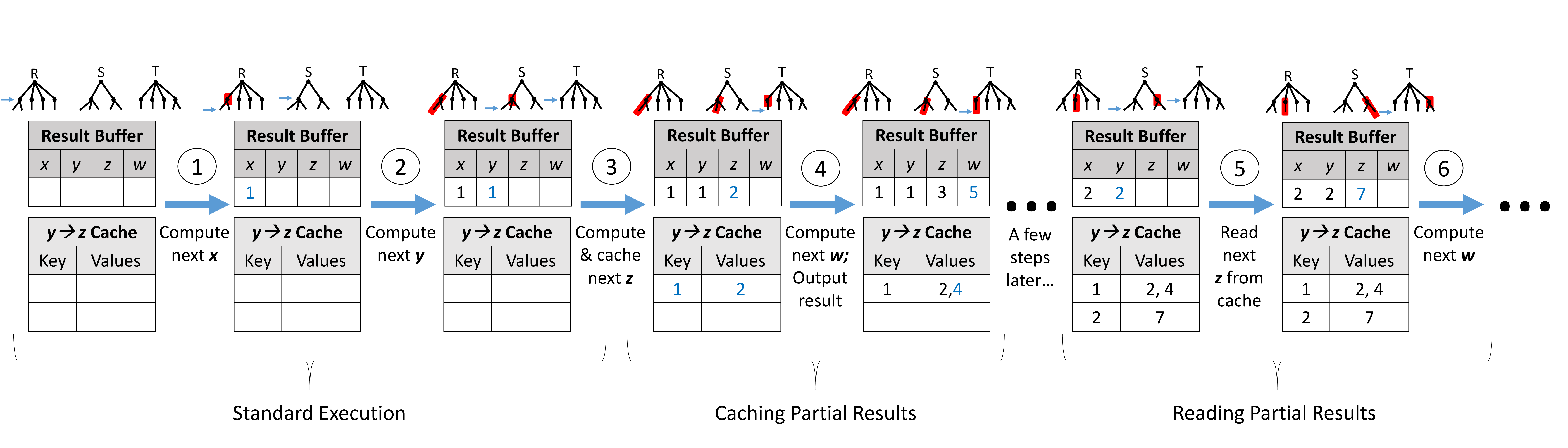}
    \caption{Cached TrieJoin execution and caching flow of an example Path-4 query. Each step marks the current partial join path on the tries (top) \label{fig:ctjflow}}
\end{figure*}

{
\begin{algseries}{t}{\label{alg:cachedtriejoin}The $\algname{Cached TrieJoin}$ algorithm}
\begin{insidealg}{CachedTrieJoin}{d, inds, cache, res}
\IF{$d=n+1$} 
\STATE \textbf{save} $res$ 
\STATE \textbf{return}
\ENDIF
\STATE $keyIDs:=\algname{cacheKeysOf(d)}$
\STATE $valIDs:=\algname{cacheValsOf(keys)}$
\IF{$res[keyIDs]$ is a cache hit in $\cache$} 
\FORALL{$cached$ in $\cache(res[keyIDs])$}
\STATE $res[valIDs]:=cached$
\STATE $\algname{AdjustTries(inds,res)}$
\STATE $next:=\max(valIDs)+1$
\STATE $\algname{CachedTrieJoin}(next, inds, cache, res)$
\ENDFOR
\STATE $\algname{ResetTries(inds, min(valIDs))}$
\STATE \textbf{return}
\ENDIF
\FORALL{matching values $a$ for attribute $d$ in $inds$}
\STATE $res[d]:=a$
\STATE $\algname{AdjustTries(inds,res)}$
\STATE $\algname{CachedTrieJoin}(d+1, inds, cache, res)$
\IF{$d=\max(valIDs)$}
\STATE $\algname{ApplyCaching(\cache,res[keyIDs],res[valIDs])}$
\ENDIF
\ENDFOR
\STATE $\algname{ResetTries(inds, d)}$
\end{insidealg}
\end{algseries}
}

\subsubsection{Indexes}
CTJ saves its relations in tries, a multi-level data structure commonly used in WCOJ systems~\cite{Aberger:2016:ERE:2882903.2915213}, column stores and graph engines~\cite{Stonebraker:2005:CCD:1083592.1083658, Hong:2012:GDE:2150976.2151013}. Section~\ref{sec:arc} elaborates on the physical layout of our indexes. Given a relation, for example the $R(x,y)$ relation with two attributes in Figure~\ref{fig:index_ex}, the trie representation is as follows:
\begin{itemize}
    \item Each attribute, such as $x$, is a level in the trie.
    \item Each unique path from the root to a leaf is an entry in the relation. For example, the path $(1,1)$ corresponds to the entry of $x=1$ and $y=1$ in $R(x,y)$.
    \item The siblings are sorted.
\end{itemize}

\subsubsection{The Cached TrieJoin algorithm}
Next, we elaborate on CTJ, one of our main building blocks. Figure~\ref{fig:ctjflow} will be used to illustrate the CTJ execution flow. The algorithm is presented in Figure~\ref{alg:cachedtriejoin}. CTJ first orders the variables (e.g, $x \rightarrow y \rightarrow z \rightarrow w$). Then, it looks for matches for each variable in turn from first ($x$) to last ($w$). Initially, the cache is empty (line 14 in Fig.~\ref{alg:cachedtriejoin}). Starting from $x$, CTJ will search all the tries that contain $x$ (e.g., $R(x,y)$) for a match (\circleCtj{1} in Fig.~\ref{fig:ctjflow}). Each match is found using a variation of merge-join, called \e{leapfrog-join}~\cite{DBLP:conf/icdt/Veldhuizen14}, that uses lowest upper bound searches to leap over the variable levels until a match is found. If a match is found, it sets the $x$ value in the result buffer (line 15 in Fig.~\ref{alg:cachedtriejoin}).


Before continuing to the next variable (e.g., $y$), CTJ adjusts the tries (Line 16) such that they align on the children of the current partial join path (the nodes below the paths marked in red in Fig.~\ref{fig:ctjflow}). For example, after step \circleCtj{1}, the $R$ trie is set on the $y$ child nodes of $R(x)=1$. Finally (Line 17), the algorithm calls CTJ to look for a match for the next variable $y$ (\circleCtj{2}). In practice, CTJ uses queues instead of recursion. Once all the variables are set (\circleCtj{4}), CTJ saves the result (line 2). If no other matches are found for the current variable, CTJ resets the adjusted tries (line 20) to focus on the previous attribute.

CTJ uses cache for partial join results to avoid the computation of recurrent partial joins. 
Given a join variable (e.g., $z$), CTJ extracts the variables that can be used as keys for caching (e.g., $y$) and the variables that are cached by these keys (lines 4--5).
During the standard execution, CTJ checks the cache if the value of the current variable is stored in the cache. If so, it uses the results from the cache and adjusts the tries accordingly (lines 9--11). For example, \circleCtj{5} finds $y=2$ in the cache and reads the value of $z$ from the cache instead of recomputing the join for $z$. Writing to the cache is done during the standard execution (\circleCtj{3}), after finding a valid match for a cached entry (line 19).




\section{\sysname Architecture}\label{sec:arc}

\begin{figure}
    \includegraphics[width=\columnwidth,trim={0ex 0ex 0ex 0ex}, clip=true]{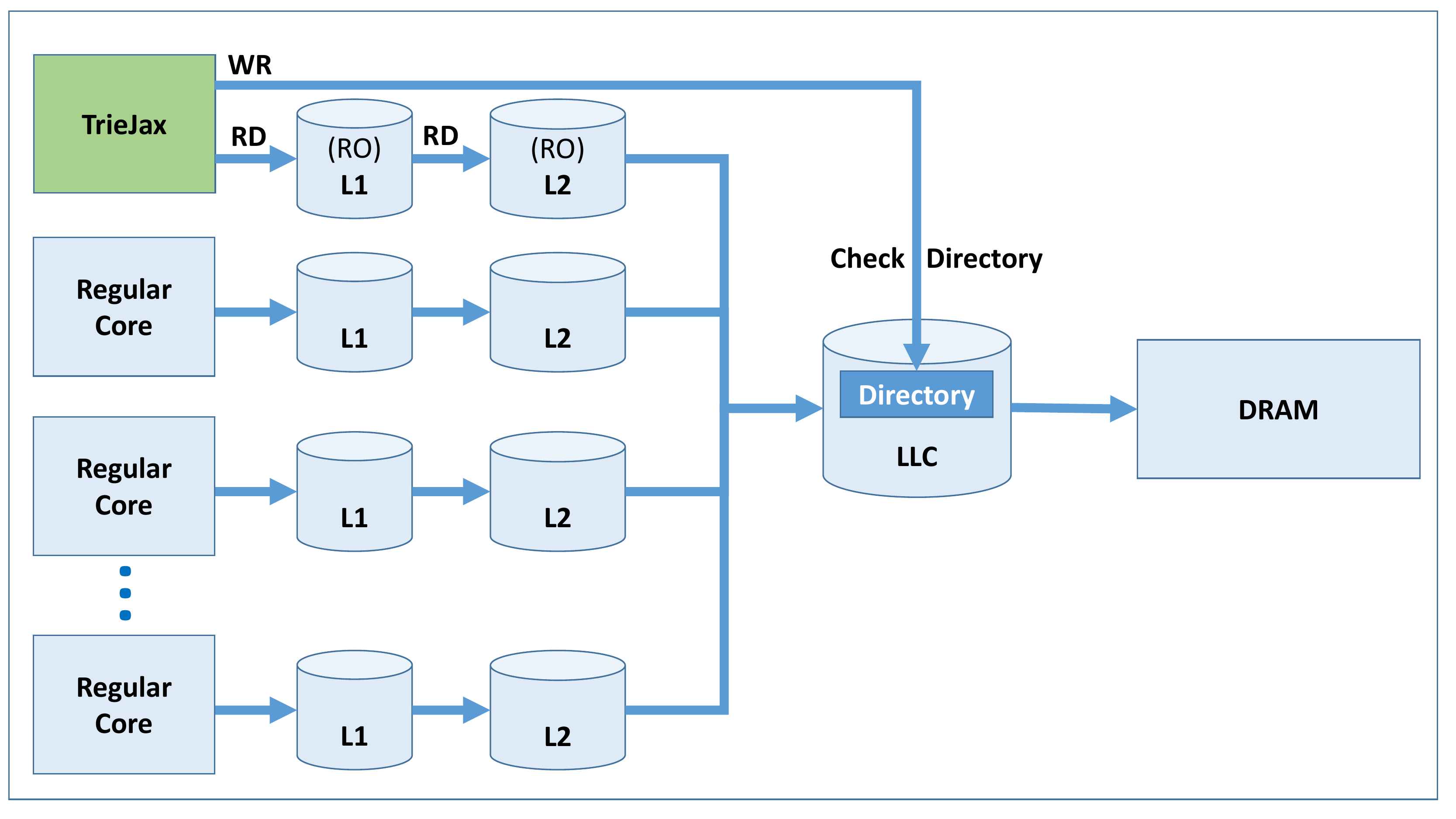}
    \caption{System architecture with \sysname as another core and the communication between \sysname and the memory \label{fig:sysArc}}
\end{figure}

\subsection{System-level overview}
\sysname is designed as a small, on-die co-processor that integrates to the main processor like an additional core, as depicted in Figure~\ref{fig:sysArc}. This design allows \sysname to use the main processor's memory system and avoids the coherence and data transfer overheads that plague discrete accelerators.

\sysname uses a local read-only L1 and L2 caches that, in turn, are used for caching the index data. Since \sysname stores its intermediate results in a private scratchpad as we describe later on, only the final results of the join operations are written to memory. 
The memory system, therefore, streams these write operations directly to memory and bypasses the private caches to avoid cache thrashing and congestion in the cache queues.
For example, on some of the benchmarks we evaluate (e.g., \e{path4} query), where the size of the resulting join table is extremely large, bypassing the private caches improves performance by up to $2.5 \times$.

\hide{
Coherency is not maintained between the CPU and \sysname. Therefore, the CPU must clear the index data from its caches before initiating the query. To avoid any coherency issues, \sysname write operations check the LLC directory before being forwarded to the DRAM, invalidating other instances of the output data in the rest of the caches.
}

The regular processor cores communicate with \sysname using a co-processor interface similar to that of ARM~\cite{ArmISA} or RISC-V~\cite{RISCvISA}. 
For instance, we use a parallel of the ARM \e{LDC} command to load the query to the \sysname internal memory.

\begin{table}

\centering
\begin{tabular}{ | l | l | }
 \hline
 Name & Query \\ 
 \hline \hline
 Path-3 & $path3(x,y,z) = R(x,y), S(y,z).$ \\ 
 \hline
 Path-4 & $path4(x,y,z,w) = R(x,y), S(y,z),T(z,w).$ \\ 
 \hline
 Cycle-3 & $cycle3(x,y,z) = R(x,y), S(y,z),T(z,x).$ \\ 
 \hline
 Cycle-4 & $cycle4(x,y,z,w) = R(x,y), S(y,z),T(z,w),$  \\ 
 &   \hspace{2.6cm} $U(w,x).$ \\ 
 \hline
 Clique-4 & $clique4(x,y,z,w) = R(x,y), S(y,z),T(z,w),$ \\
  &   \hspace{2.7cm} $U(w,x),V(z,x),W(w,y).$ \\ 
 \hline
\end{tabular}
\caption{Graph pattern matching operations used to evaluate \sysname and their mapping to join queries (shown in datalog format, for brevity).
\label{fig:queries}}  
\end{table}

\subsection{Query language and index memory layout}
\begin{figure}
    \centering
    \includegraphics[width=0.9\columnwidth,trim={40ex 15ex 40ex 10ex}, clip=true]{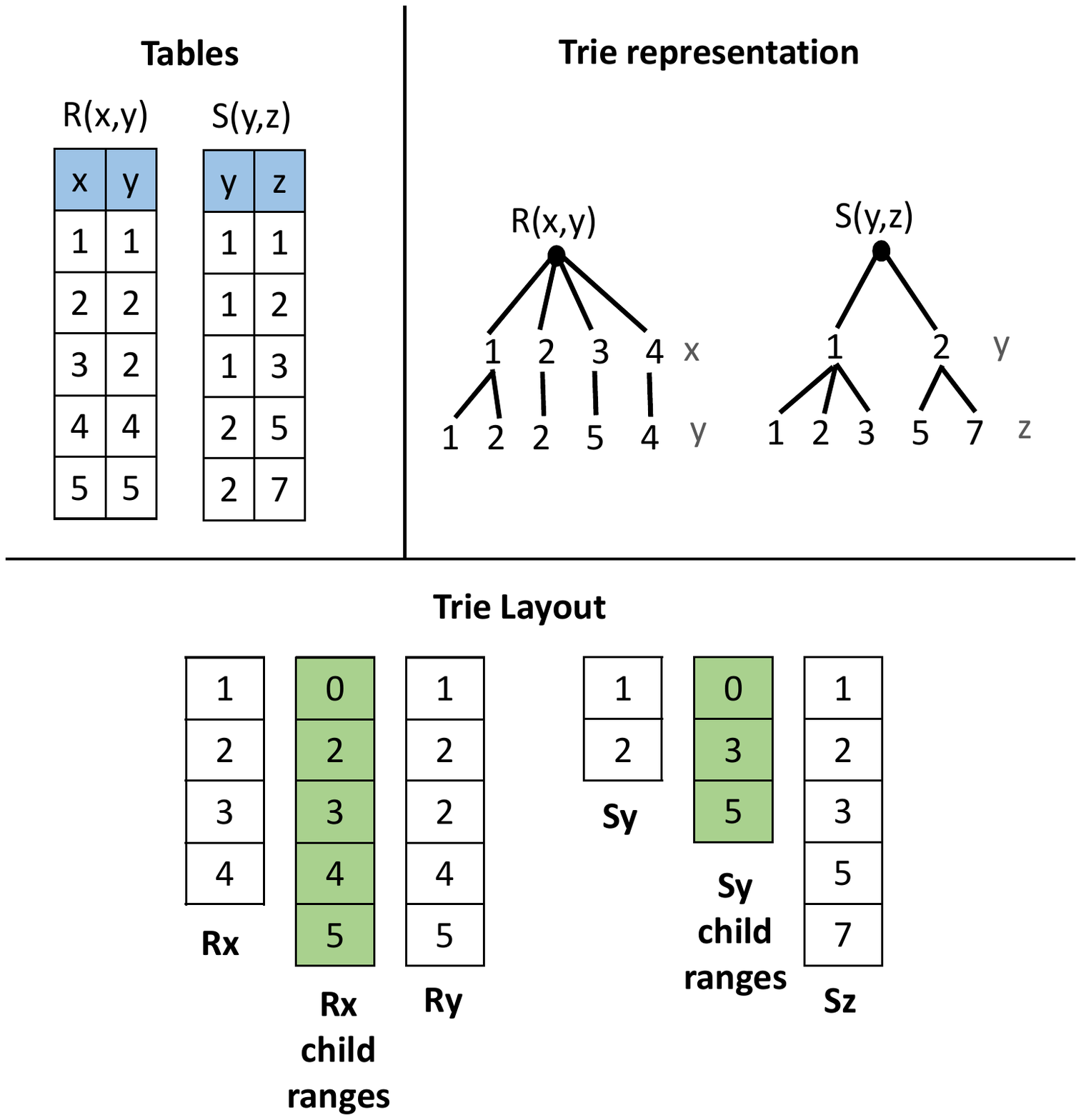}
    \caption{Tables for an example query $path3(x,y,z) = R(x,y), S(y,z).$ (top left), their trie representations (top right), and their memory layout (bottom) \label{fig:trieExmp}
    }
\end{figure}

We use the CTJ compiler~\cite{DBLP:conf/edbt/KalinskyEK17} to compile SQL join queries for \sysname. 
Table~\ref{fig:queries} lists the graph pattern matching queries used in our evaluation (Section~\ref{sec:eval}) and their mapping to join queries (for brevity, the table uses the compact datalog format rather than SQL).

Figure~\ref{fig:trieExmp} illustrates a trie layout in \sysname.
\sysname uses in-memory trie indexes (described in Section~\ref{subsec:ctj}) in a physical layout similar to that of EmptyHeaded~\cite{Aberger:2016:ERE:2882903.2915213}. 
Specifically, this layout stores the unique values of the first join attribute in the relation as a sequential array. The next join attribute is then stored by concatenating the values that match the previous attribute to a continuous array. To identify which values of the second join attribute belong to which elements in the previous attribute, the child ranges array lists the corresponding ranges of the second attribute.
For example, Figure~\ref{fig:trieExmp} shows that the values $\{(1,1), (1,2)\}$ in $R$ can be extracted by focusing on $R(x=1)$, extracting the ranges of its children from the respective indexes $[0,2)$ in the \emph{Rx child ranges} array, and accessing the extracted index range in the $Ry$ array.





\begin{figure*}
    \includegraphics[width=\textwidth,trim={0ex 26ex 0ex 25ex}, clip=true]{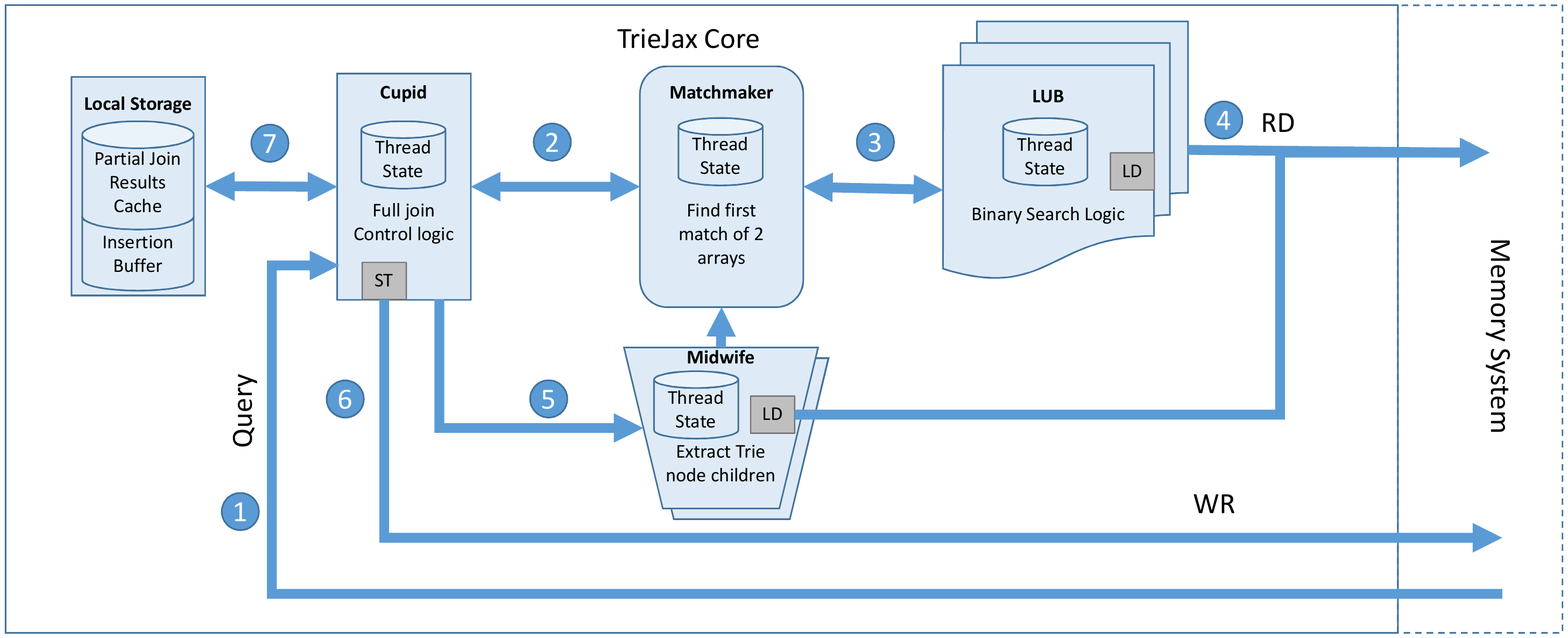}
    \caption{\sysname core components and its high-level operational flow. \label{fig:highlevel}}
\end{figure*}

\subsection{The \sysname operational flow}
Figure~\ref{fig:highlevel} presents the high-level design of \sysname and the communication between the different components it utilizes to answer graph pattern matching queries. The five major building blocks are lowest upper bound (\lub), \mmaker, \midwife, \cupid and the \irs (\tirs). We illustrate the operational flow of \sysname on the \e{Path3} example query in \fig{fig:trieExmp}. 
We begin by describing the operational flow as single threaded. 
\sect{sec:mt} then describes how \sysname incorporates multi-threading internally to hide memory latency.
The micro-architecture and the internal flow of each component are described in \sect{subsec:blocks}.

The query execution begins by loading the compiled query to a local read-only store in \cupid, the component that controls the execution of the full join query (marked \step{1} in \fig{fig:highlevel}). \cupid starts with the first join variable $x$, by extracting the pointers to the trie arrays of $x$ embedded in the compiled query structure (e.g., $Rx$), and sending them to \mmaker (marked \step{2} in \fig{fig:highlevel}). Until a response is returned, \cupid saves the current state (e.g., result buffer, parameter id) in its local State Store.

\mmaker is in charge of finding the first matched value of $x$ in all the tries (based on the Leapfrog Join algorithm described in Section~\ref{subsec:ctj}). 
Our running example (\fig{fig:trieExmp}) only has one array ($Rx$), so \mmaker will request the \lub unit to find the first value for $x$ in $Rx$ (\step{3}). 
\lub performs a binary search to look up the lowest upper bound of a value in a given array.
The unit will search for the index in the trie array in which the value is stored (\step{4}), e.g., $Rx_{index}=0$, and send the result back to \mmaker. \mmaker will send the read value and index (e.g., $x=0$, $Rx_{index}=0$) back to \cupid, which will save the index in the current state of variable $x$.
Similarly to \cupid, \mmaker and \lub save their local state in their State Store when sending a request and reads the data when receiving the response.


Next, \cupid will continue to the next variable $y$ and look for a match for $y$. It will read the $Ry$ and $Sy$ array pointers from the compiled query. As described above, the $Ry$ array stores \e{all} the values of the child nodes of $Rx$. To extract the $Ry$ range that belongs to the current $x$ value, \cupid will send the $x$ index, the $Ry$ pointer, and the \e{$Rx$ child ranges} pointer to \midwife (\step{6}). The \midwife component is in charge of extracting the children of a node in the trie.
For example, to extract the $Ry$ values for $Rx_{index}=0$, \midwife will read from the child ranges array the start and end ranges from indexes 0 and 1, respectively. The final $Ry$ range (e.g., $Ry[0:2]$) will be sent to \mmaker.

Once \mmaker receives both $Sy$ and $Ry$ array ranges sent from \cupid and \midwife, it will look for the first match of the two array ranges. 
It will send the pointer of the first value in $Ry$ and the $Sy$ array pointer to \lub (\step{3}). \lub loads the value of $Ry$ from memory and uses binary search to look for its lowest upper bound in $Sy$ (e.g., $Sy=1$). The result is returned to \mmaker, which extracts its state from the State Store and checks if a match $LdVal$ was found.
If so, it will return the data to \cupid that, in turn, will set the state for variable $y$ and continue to the next variable $z$.
If not, it will use \lub to look for $LdVal$ in the $Ry$ range, using iterative LUB searches on $Sy$ and $Ry$ until a match is found or it reaches the end of an array. If no match is found, \mmaker will return a failed response to \cupid, which will restore the previous variable and look for its next match. Once a match is found for $x$, $y$ and $z$, \cupid will write the result to memory. \sysname uses a small write buffer and sends results to main memory when the buffered results exceed the size of a cache line.

\begin{figure}[t]
    \includegraphics[width=\columnwidth,trim={42ex 20ex 62ex 14ex}, clip=true]{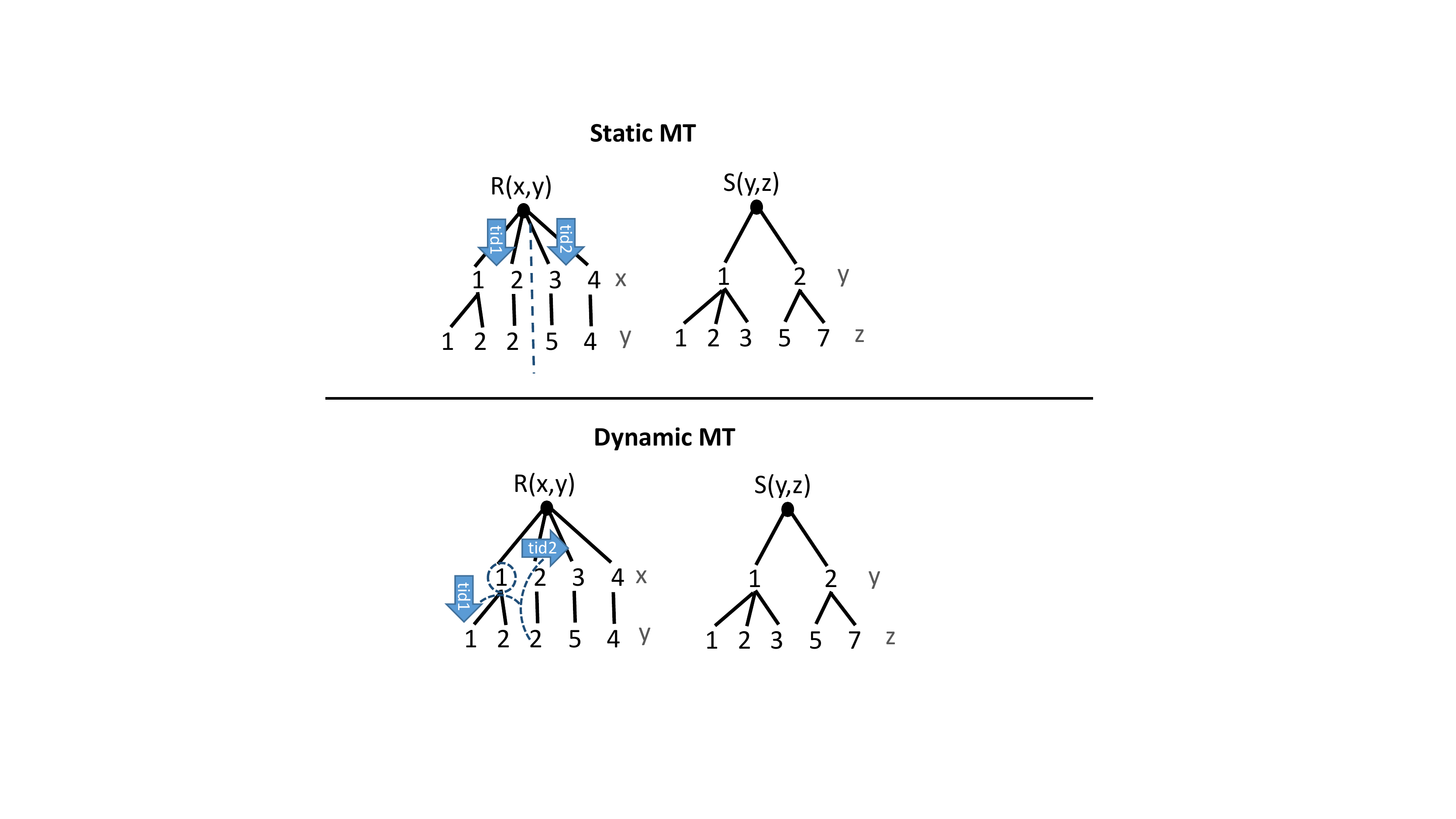}
    \caption{\sysname MT schemes for the path--3 query. Static MT (top) divides the query search space statically based on the first join attribute. Dynamic MT (bottom) can spawn a new thread on every match to continue with the query while the current thread focuses on the sub-query given the match.  \label{fig:mt}}
\end{figure}


\subsection{Multi-threading in \sysname}
\label{sec:mt}
\sysname's use of multi-threading (MT) achieves two performance benefits. 
First, MT parallelizes the trie join and thereby extracts memory-level parallelism (MLP), which allows \sysname to hide memory latency. 
Second, MT pipelines the operations on the \irs, and enables \cupid to lookup pre-computed partial results while the rest of the accelerator is looking for new results.

Figure~\ref{fig:mt} presents two different MT schemes, static and dynamic. 
The static MT scheme divides the arrays of the first attribute between the different threads. This scheme is similar to software MT solution in DBMS systems such as EmptyHeaded~\cite{Aberger:2016:ERE:2882903.2915213} and GPU query engines~\cite{DBLP:conf/vldb/2014adms}. The disadvantage of this scheme is an unbalanced workload distribution between the different \sysname threads. For example, in Figure~\ref{fig:mt}, \e{tid1} will generate many results while \e{tid2} will finish quickly without any results.

Dynamic MT, on the other hand, balances the load by dynamically allocating new threads on each match. On a match, the \cupid unit splits the search space to two sub-spaces. 
The original thread is then bound to the search space containing the current partial result, and the new thread will be bound to the search space after the current match. 
For example, in Figure~\ref{fig:mt}, a match is found for $x=1$ while using dynamic MT. After the match, \e{tid1} will compute the results for $x=1$ and \e{tid2} will compute the results for $x>1$.

On its own, however, dynamic MT might incur a slow initialization time for queries with infrequent matches due to fewer matching opportunities. \sysname thus combines both static and dynamic MT.

Each component supports multi-threading by replicating its internal execution state.
As shown in Figure~\ref{fig:highlevel}, each component in the system maintains a small local memory to store the execution state when  waiting for a response from another component (e.g. when \cupid requests \mmaker to find the next match for a variable). 
Replicating this state allows each component to maintain multiple operations in-flight.
For duplicated components, such as \midwife, we use banked stores to support parallel accesses.
\sect{subsec:eval_mt} examines the performance impact of MT and identifies the number of threads required to achieve performance/storage balance.

\subsection{Caching with threads}
Locality across partial join results enables \sysname to break query execution into two parallel flows using the \sysname\xspace \irs (\tirs). 
In the main flow, \cupid, \midwife, \mmaker and \lub construct new partial results.
In the other flow, \cupid maintains partial results in the \tirs and checks if the stored partial results can be used instead of executing the main flow and recompute them.
Decoupling the flows and using MT in both allows the units to execute concurrently.

The \tirs stores intermediate partial join results based on the cache structure provided by the \sysname compiler, namely which attributes are keys and their corresponding cached attributes. In our caching example from Section~\ref{subsec:ctj} (\fig{fig:ctjflow}), the cache entry of \tirs stores the partial join values and indexes of $z$ (e.g., the values $2$ and $4$ and their indexes) and it is keyed by a hash of the corresponding $y$ value (e.g., $1$). The indexes in the trie array are stored to allow the expansion of the children nodes by \midwife.

When \cupid finds a match on a key attribute (e.g., $y$), it searches \tirs for its partial join results. For instance, the values of $z$ for the given $y$ (similarly to step \circleCtj{5} in Figure~\ref{fig:ctjflow}). 
In the case of a cache hit, it uses the cached results instead of recomputing them. 
Otherwise, \tirs will allocate a new entry for the array of intermediate results, and \cupid will use the main flow to compute the values and set them (and their indexes) in the entry. 

Since partial join results can have variable lengths, each entry contains the number of values currently stored in the entry. If an entry overflows, when it attempts to store more intermediate results than the entry size, it is deallocated to avoid storing incomplete results.




A major challenge when using MT to fill the \tirs is managing read/write race conditions. We solve this issue by adding an insertion buffer that stores entries that were not fully analyzed, meaning that \sysname did not finish analyzing all the paths under the entry key. Once an entry is fully analyzed it is copied atomically to the \tirs. 
Another MT challenge is write/write races. A keen reader will notice that the same partial join result can be accessed from different paths. If a partial join result is available in the cache, it will be used by the querying thread. However, if the entry is still in the insertion buffer, two threads from different paths might try to append values to the same entry. To avoid this race, the insertion cache uses all the values leading to the key to validate that the values are stored from just one path. 
For example, when looking on the dynamic MT in Figure~\ref{fig:mt}, we can see that \e{tid1} computes the results for $(x=1,y=2)$, while \e{tid2} looks for the results of $(x=2,y=2)$. If the results for $y=2$ are not in the \tirs, \e{tid2} will compute the results and store them in the insertion buffer. While $y=2$ is in the insertion buffer, the value of $x$ for \e{tid2} is different and the \e{tid2} results will not be stored.

The last caching challenge is how to determine that an entry is fully analyzed. With dynamic MT, multiple threads can work on the same cache entry in parallel. New threads can even be created on the fly to help analyze the cache entry.
To avoid race conditions we add a thread counter to each entry that maintains the number of threads currently working on the entry. Each thread that is involved with a cached entry notifies the cache of its allocation or deallocation to update the count. Once the count reaches zero, the cached entry analysis is done and the entry is copied to the \tirs.

\subsection{The \sysname components}
\label{subsec:blocks}
We now turn to describe the microarchitecture and internal flow in each \sysname component.

\begin{figure}
    \includegraphics[width=\columnwidth,trim={2ex 40ex 0ex 20ex}, clip=true]{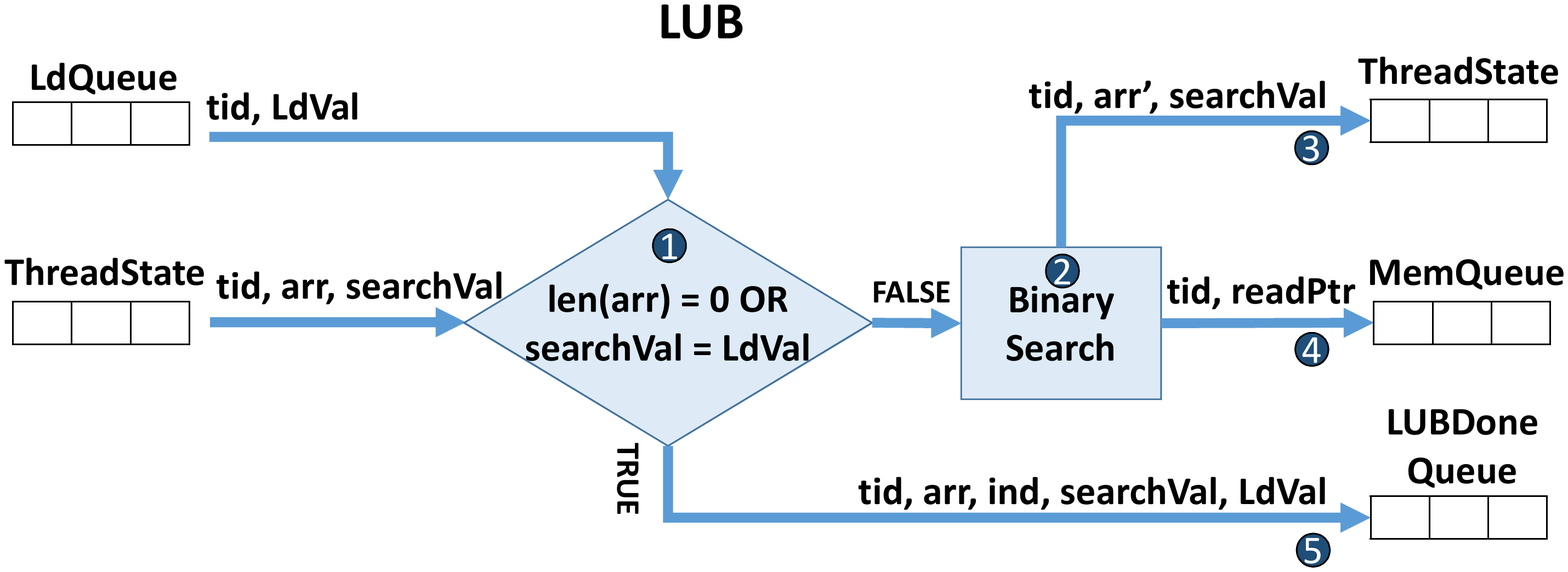}
    \caption{The logical flow of \lub: \circleLub{1} If array is empty or match found return current result. \circleLub{2} Otherwise, Update array range (namely \e{arr'}), read middle of array from memory (\circleLub{4}) and store current state in local Thread Store (\circleLub{3}).  
    \label{fig:lub}}
\end{figure}

\mparagraph{\lub} This component searches for a value in a sorted array using binary search. If the value does not exist, it returns the lowest upper bound. This component has a load (LD) unit to communicate with the memory.
\fig{fig:lub} presents the logical flow of \lub. 
Since most of the read operations issued by \sysname are part of binary search operations, encapsulating the binary search logic in the specialized \lub component allows us to duplicate the component and generate multiple memory accesses that look for matches on different sub-arrays concurrently.

\begin{figure}[t]
    \includegraphics[width=\columnwidth,trim={0ex 30ex 0ex 20ex}, clip=true]{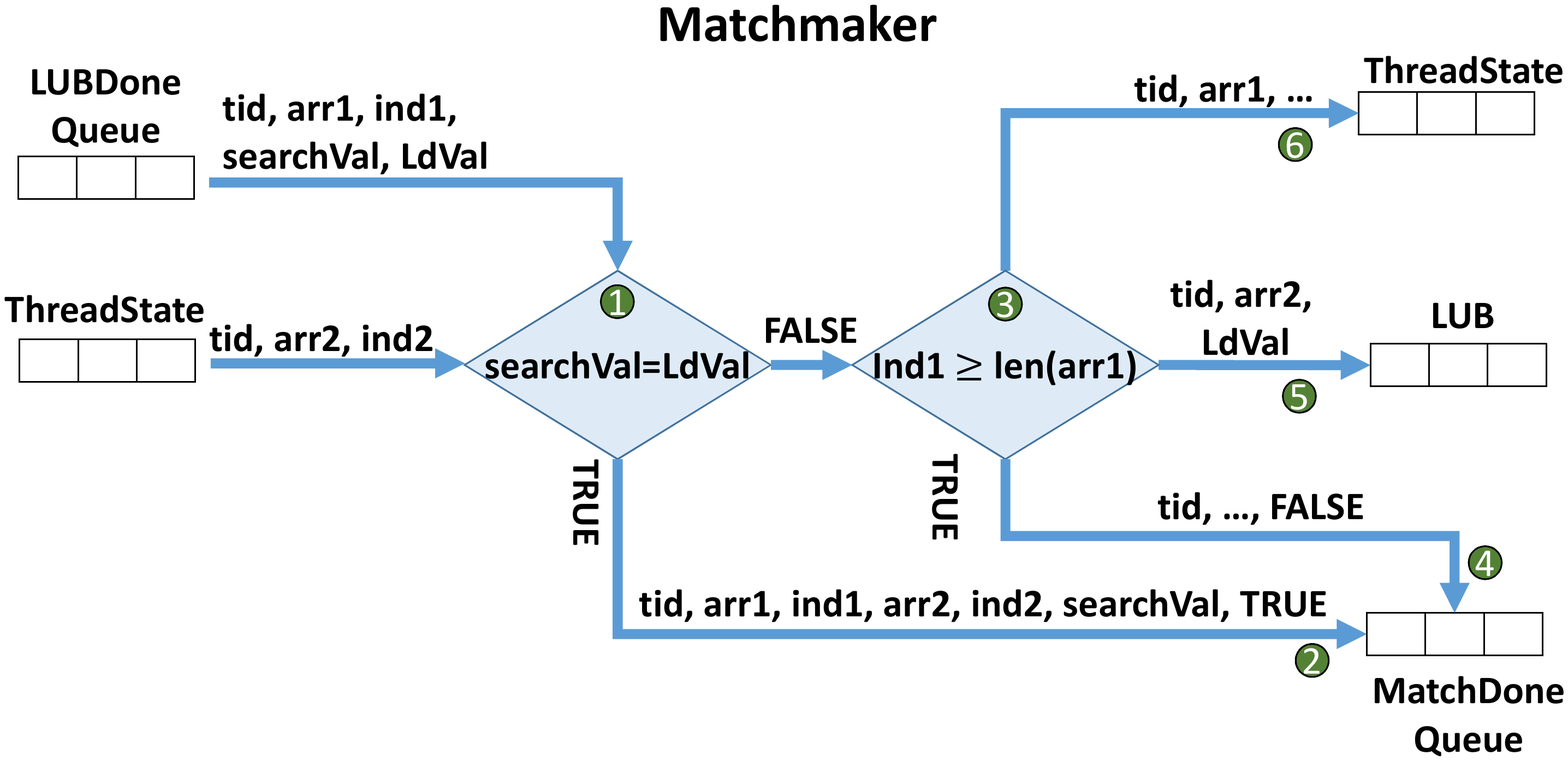}
    \caption{The logical flow of \mmaker: \circleMak{1} If \e{match} found return True (\circleMak{2}). \circleMak{3} Otherwise, check if \e{index1} in range. If not, return False (\circleMak{4}). Otherwise, look for lowest upper bound in \e{arr2} (\circleMak{5}) and store thread state in the local Thread Store (\circleMak{6}). \label{fig:mmaker}}
\end{figure}

\mparagraph{\mmaker} This component implements the leapfrog join algorithm described in \sect{subsec:ctj} for one join variable. \fig{fig:mmaker} presents the logical flow of \mmaker and the queues used to communicate with \lub and \cupid. 
Given two array ranges, this unit communicates with \lub to find the first value that is contained in both arrays. If no such value is found, the component returns false.

\begin{figure}[t]
    \includegraphics[width=\columnwidth,trim={0ex 35ex 0ex 22ex}, clip=true]{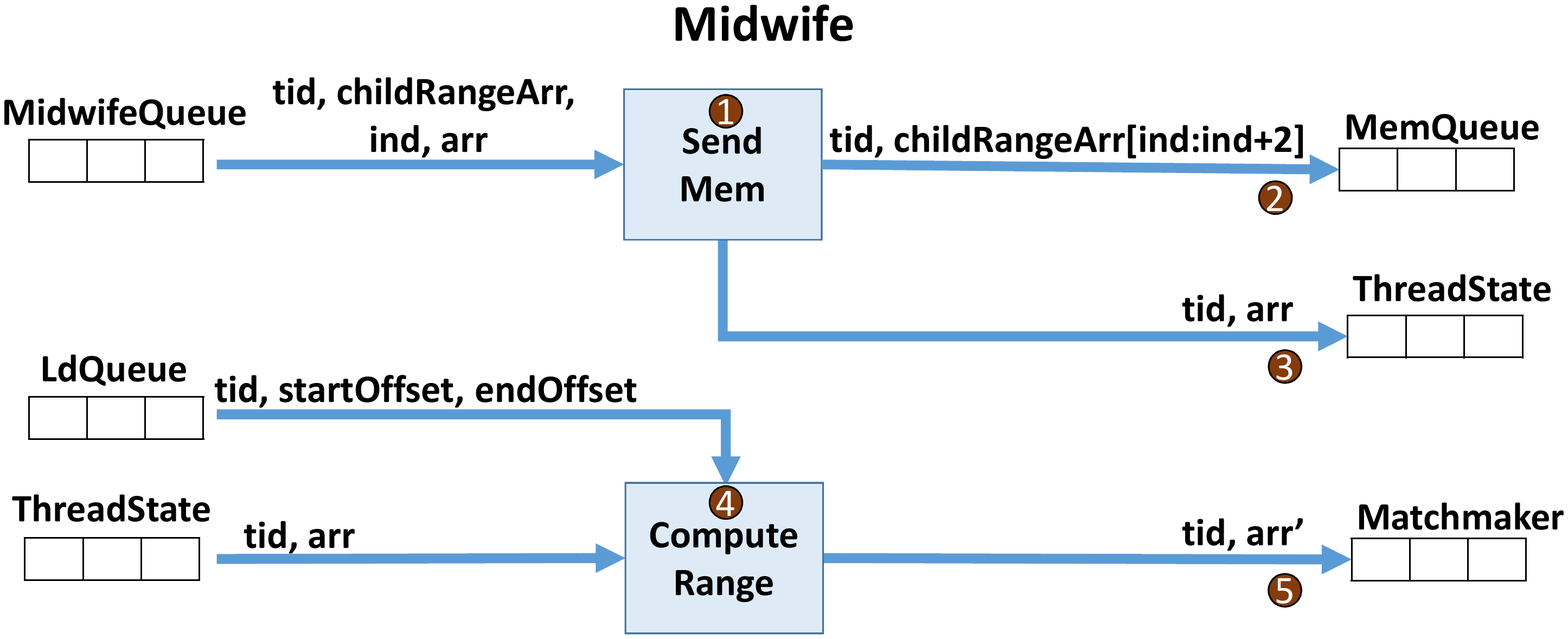}
    \caption{The logical flow of \midwife: \circleMid{1} Midwife receives the child range array of a parameter in the trie and (matched) value index. \circleMid{2} It reads from memory the start and end offsets of the next parameter array that fits the (matched) value. Next, \circleMid{4} it computes the next parameter array range and sends it to \mmaker (\circleMid{5}). 
    \label{fig:midwife}}
\end{figure}

\vskip0.5em

\mparagraph{\midwife} This component is used to extract the child nodes of a value node in the trie data structure, following the logic presented in \fig{fig:midwife}. The component computes the  memory range in which the child nodes of a given parent node are stored. 
The inputs to the unit are a pointer to a child node array and a parent index. 
For example, given the \e{Rx child ranges} array in \fig{fig:trieExmp} and index $0$ of \e{Rx}, \midwife extracts the range $[0,2)$ and returns the $Ry[0:2]$ range (start and end pointers) to \mmaker. The LD unit is used to access the children array. This component is duplicated to handle up to two child node ranges in parallel.

\begin{figure}
    \includegraphics[width=\columnwidth,trim={11ex 10ex 10ex 10ex}, clip=true]{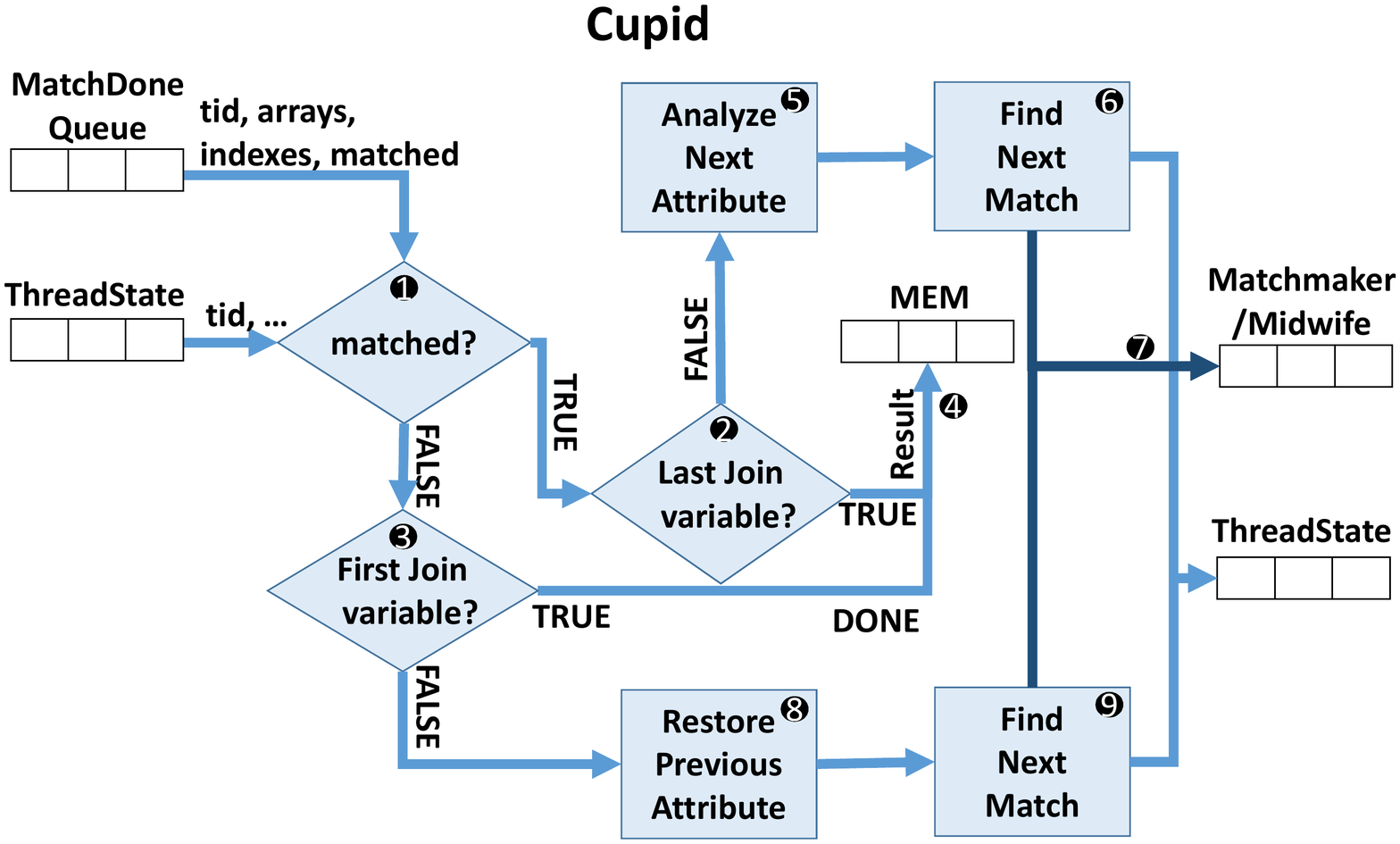}
    \caption{The logical flow of \cupid (single thread, no cache): \circleCup{1} If a match is found for the current join variable, \circleCup{2} check if it is the last join variable. If so, save the result to memory (\circleCup{4}). If not, save the arrays, matched indexes and value of the current join variable, and extract the indexes of the next join attribute from the query structure (\circleCup{5}). If the join variable is in the first trie level send it directly to \mmaker, otherwise use \midwife to extract the parameter values (\circleCup{6}). If a match was not found, check if it is the first join variable (\circleCup{3}). If so, \circleCup{4} write a \e{DONE} token to memory. Otherwise, \circleCup{8} restore the previous join variable data and \circleCup{9} look for the next match similarly to \circleCup{6} \label{fig:cupid}}
\end{figure}

\vskip0.5em

\mparagraph{\cupid} This unit manages the execution of the full join query. Figure~\ref{fig:cupid} shows the single threaded logical flow of \cupid. For each join attribute $k$, it uses the \mmaker to look for the next match in all the indexes that contain $k$. If a match is found, it updates the result buffer and moves to the next join attribute. Otherwise, it backtracks and looks for the next match for the previous join attribute. Finally, the unit uses its store (ST) unit to save the final results to memory. This component is also in charge of the threads management and partial join results caching. More information can be found later in this section.



\subsection{\sysname local memory system}
There are three local memory stores in \sysname. The first is the \tirs, which is accessed by \cupid. The second is a constant read-only memory that stores the query structure and cache structure described in \sect{sec:arc}. Finally, each component holds a small local memory for maintaining the thread state. 
For example, \lub units store the searched value and array range before sending the request to memory. Once the request is returned, the thread information is read from the \lub thread store and continues according to the result. 

We use SRAM for the local memory stores. The \tirs is the biggest store, amounting to 4 MB (for brevity, we do not present the full design space exploration for the cache size). This store uses 4 banks to allow fast concurrent accesses. The \cupid thread store is the second biggest amounting to 16 KB, while the remaining stores containing less than 512 B. It supports 32 threads, a configuration which offers the best performance/storage as examined in \sect{subsec:eval_mt}.

\section{\sysname Evaluation} \label{sec:eval}


In this section, we present our evaluation of \sysname when executing graph pattern queries and compare its performance and power benefits to four state-of-the-art baselines: Cached Trie Join (CTJ)~\cite{DBLP:conf/edbt/KalinskyEK17} and EmptyHeaded~\cite{Aberger:2016:ERE:2882903.2915213} are software systems, and Graphicionado~\cite{Ham:2016:GHE:3195638.3195707} and Q100~\cite{Wu:2014:QAD:2654822.2541961} are hardware accelerators.
Our evaluation focuses on the following core questions:

\begin{itemize}
    \item How does the performance of \sysname compare to the baselines?
    \item What is the power consumption of \sysname compared to the baselines?
    \item What is the effect of the \sysname multi-threading on performance?
    \item How the \irs affect the performance?
\end{itemize}

\subsection{Methodology}

\partitle{\sysname}
We implemented all the \sysname building blocks using  PyRTL~\cite{DBLP:conf/fpl/ClowTDGMS17}. 
We then used Cadence Innovus and Design Vision in tandem with the OpenCell 45nm design library to synthesize and place\&route the  Verilog code generated by PyRTL. 
The design achieves a fixed frequency of 2.38GHz (critical path of 0.42ns), and the results are for dynamic multi-threading with 32 threads (unless otherwise noted).

The timing and power figures obtained by the physical design were used to drive a cycle-accurate simulator of \sysname. The simulator models all micro-architectural components described in Section~\ref{sec:arc}.
We integrated the simulator with Ramulator~\cite{DBLP:journals/cal/KimYM16} to obtain accurate performance of the memory system. 
The DRAM energy is simulated with DRAMPower~\cite{chandrasekar2012drampower} using the memory traces from Ramulator. 
The performance and power of the on-chip SRAM memory were simulated using Cacti 6.5~\cite{cacti}. 
The \sysname uses a default \tirs size of 4 MB, which includes the insertion buffer. The off-chip memory is simulated as 64GB of DDR3\_1600 DRAM with two channels. The total \sysname core area is $5.31 mm^2$.

\partitle{Baselines}
We compare \sysname to the Q100 and Graphicionado hardware accelerators and to CTJ and Emptyheaded software systems (these baselines are discussed in detail in \sect{subsec:related}). 
Since we do not have the original code for the hardware accelerators, we estimate the performance of Q100 and Graphicionado based on their original software baselines, MonetDB and GraphMat, respectively. 
We configured the original software baselines as described in the original Q100 and Graphicionado papers and scale the baselines' results given the best speedup and energy improvement reported in the papers. 
We believe this methodology provides a comparison that is favorable to both Q100 and Graphicionado.

In addition, we use CTJ and EmptyHeaded as highly tuned WCOJ software baselines. To achieve a comparison that is favorable to the baselines, each system is run 3 times and the minimum value is reported.

To evaluate the power consumption of all software systems we measured the power of the Package and DRAM using the Intel Running Average Power Limit (RAPL)  meters~\cite{DBLP:conf/islped/DavidGHKL10}. We sample the energy consumption during the benchmark and deduct idle energy consumption measured on the same machine for the same amount of time. 
For Q100 and Graphicionado, whose original papers did not report DRAM energy consumption, we estimate the memory energy consumption by dividing the DRAM energy consumption of the respective baseline by the accelerator speedup. 
Because the accelerators use similar algorithms to their baselines, our scaling reduces the idle memory consumption of the DRAM avoided by the speedup. 

Finally, our experimental platform is a Supermicro 2028R-E1CR24N server. The server is configured with two Intel Xeon E5-2630 v3 processors running at 2.4 GHz, 64GB of DDR3 DRAM with two channels, and is running a stock Ubuntu 16.04 Linux.

\begin{table}[t]
\small
	\centering
	\begin{tabular}{l|r|r|l}
		\toprule
		Dataset & \#Nodes &  \#Edges & Category \\
		\midrule
		ca-GrQc (grqc)                              &		5,242		&		14,496		& Collabor. \\
		soc-sign-bitcoin-alpha (bitcoin)	& 3,783 & 24,186  & Bitcoin \\
		p2p-Gnutella04 (gnu04)  &		10,876		&		39,994		& P2P \\
		ego-Facebook (facebook)	 &		4,039		&		88,234		& Social \\
		wiki-Vote (wiki)       &		7,115		&		103,689		& Social \\
		p2p-Gnutella31 (gnu31)   &		62,586		&		147,892		& P2P \\

		\bottomrule
		
	\end{tabular}
	\caption{Dataset statistics \label{table:datasets}}

\end{table}

\begin{table}[t]
\small
	\centering
	\scriptsize
	\begin{tabular}{C{1.3cm}|C{3.2cm}|C{2.9cm}}
		\toprule
		~ & TrieJax & Software frameworks \\
		\midrule
		Processing Unit & \makecell{TrieJax core @ 2.38GHz\\PRJ 4MB SRAM} & 16 $\times$ Xeon E5-2630 v3 cores @ 2.4GHz \\
		On-chip memory & \makecell{L1D ReadOnly 32KB 8-way\\L2 ReadOnly 32KB 8-way\\L3 20MB} & \makecell{L1I/L1D 32KB/core 8-way\\L2 512KB/core 8-way\\L3 40MB} \\
		Off-chip memory & \makecell{4 $\times$ DDR3-1600\\2 $\times$ 12.8GB/s channels} & \makecell{4 $\times$ DDR3-2133\\2 $\times$ 17GB/s channels} \\
		\bottomrule
		
	\end{tabular}
	\caption{Experimental configuration for TrieJax and the software baselines \label{table:configuration}}

\end{table}

\begin{figure*}[t]
    \includegraphics[width=\textwidth, clip=true]{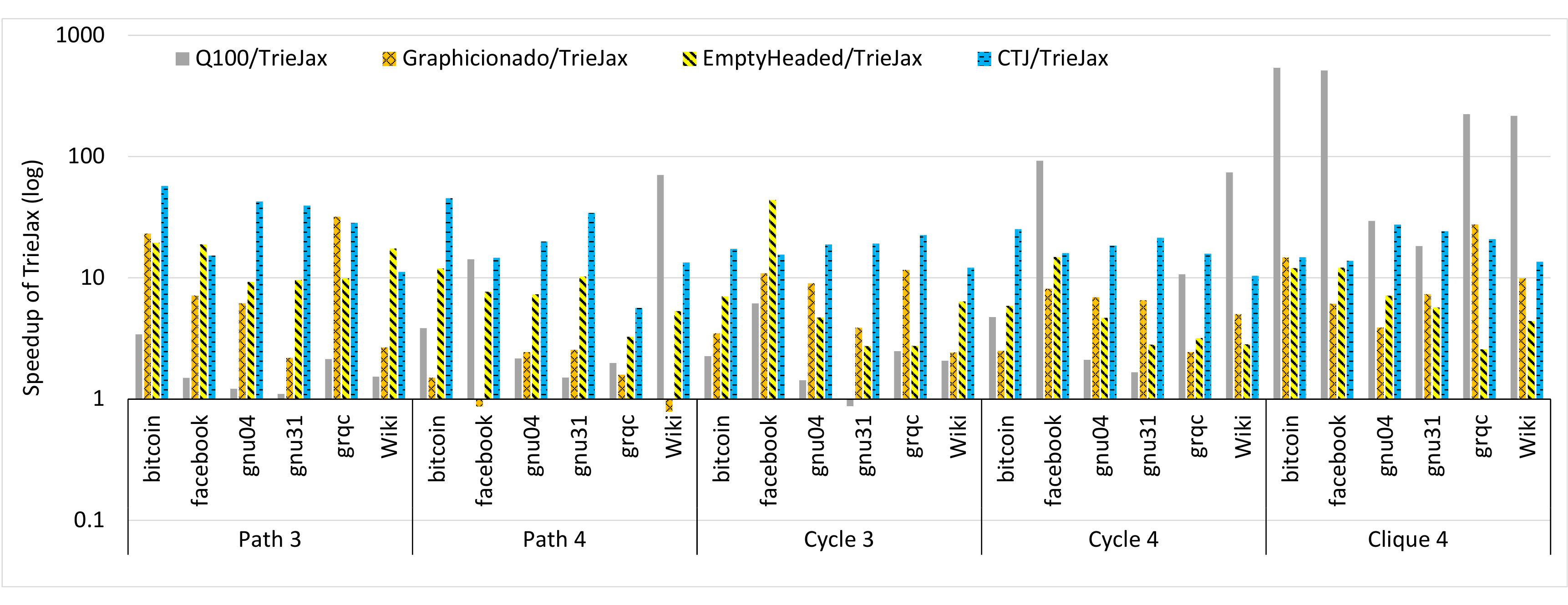}\vspace{-4ex}
    \caption{\sysname performance speedup compared to the baselines.  \label{fig:perf}}
\end{figure*}

\begin{figure}[t]
    \includegraphics[width=\columnwidth, trim={11ex 15ex 11ex 15ex}, clip=true]{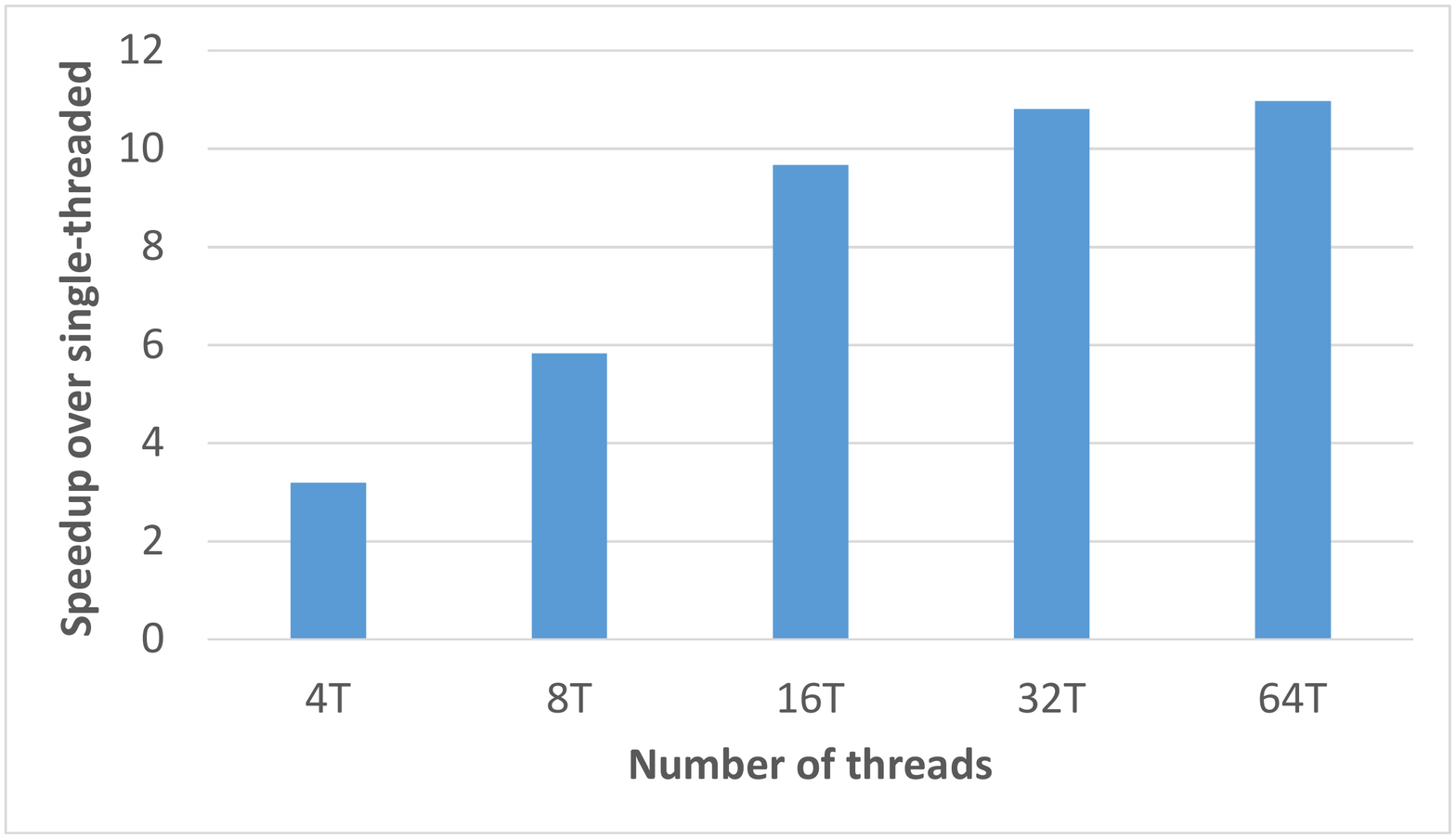}
    \caption{Performance speedup of \sysname limited to different number of dynamic threads compared to single-threaded \sysname \label{fig:mtSpeedup}}
\end{figure}

\partitle{Datasets and queries}
We test five query patterns (listed in Table~\ref{fig:queries}) on 6 different datasets. We focus on common graph pattern matching queries. Testing on linear algorithms returned comparable results to the baselines, which follows the worst-case complexity of the underlying algorithm. We leave the integration of other operators, such as SORT, to future work which can be done similarly to Q100.
Our datasets are real-world graphs taken from Stanford SNAP~\cite{snapnets}. Table~\ref{table:datasets} presents the different datasets. Due to the polynomial complexity of our queries, the runtime of the join query drastically increases on bigger datasets. Therefore, we only use datasets with simulation time smaller than five days.

\subsection{Impact of multithreading on performance \label{subsec:eval_mt}}

Figure~\ref{fig:mtSpeedup} how using different number of internal threads affect its performance.
By using only 8 internal threads, \sysname's performance is improved by $5.8 \times$, on average, compared to single-threaded implementation.
Similarly, running 32 internal threads improves the average performance speedup by $10.8 \times$ over single thread version. 
Using 64 threads, however, has a minor effect on the performance. 
We therefore choose to use 32 internal threads in our benchmarks.

\subsection{Performance comparison}
\fig{fig:perf} shows the speedup of \sysname compared to the four baselines (note the logarithmic scale) for the queries listed in \tab{fig:queries} and datasets listed in \tab{table:datasets}.

\sysname consistently outperforms the software baselines.
Compared to CTJ, \sysname achieves a speedup of $5.5 - 45 \times$ and $20 \times$ on average. 
In comparison to EmptyHeaded, which uses a highly parallel WCOJ algorithm with SIMD operations, \sysname reaches a $2.5-44 \times$ speedup and $9 \times$ on average. 

Notably, \sysname also delivers substantial speedups compared to the hardware accelerators.
\sysname is $7 \times$ faster than Graphicionado, on average, ranging between $0.8-32 \times$. The speedup is because \sysname avoids a large number of unnecessary intermediate results, that take the form of messages being passed between the different graph nodes in Graphicionado.

While \sysname offers a considerable speedup on average, Graphicionado was able to perform faster on the \e{Path4} wiki and \e{Path4} Facebook queries. In these queries, Graphicionado outperforms \sysname by up to $1.25 \times$.
The reason for this minor slowdown is that these queries generate a large number of results, and the \sysname memory system becomes a bottleneck. Nevertheless, these slowdowns may be artifacts of our optimistic estimation of Graphicionado, which does not limit its memory bandwidth.

Finally, \sysname outperforms Q100 by $63 \times$, on average, ranging from $0.9-539 \times$. This is mostly due to the inherent inefficiency of the join algorithm of Q100 that generates a large number of intermediate results. While the Q100 performance on the \e{Path3} query is comparable to \sysname for most datasets, \sysname outperforms Q100 by up to $539 \times$ on complex queries such as \e{Clique4}. Q100 is also outperformed by Graphicionado, which is aimed for graph operations and offer better parallelism and sharing of data than Q100 for large queries such as \e{Cycle4} and \e{Clique4}.

To summarize, \sysname delivers dramatic speedups over both software and hardware baselines. Thanks to its internal design that reduces the number of intermediate results and , \sysname is able to serve most of its data from its fast SRAM caches and minimizes its DRAM accesses. Finally, \sysname's aggressive use of multithreading allows it to hide the latency incurred by accessing both internal and external memories.

\subsection{Energy efficiency}

In order to understand the potential energy efficiency of \sysname, we first explore how the accelerator's energy consumption is distributed among its components.

\begin{figure}[t]
    \includegraphics[width=\columnwidth,trim={18ex 15ex 10ex 15ex}, clip=true]{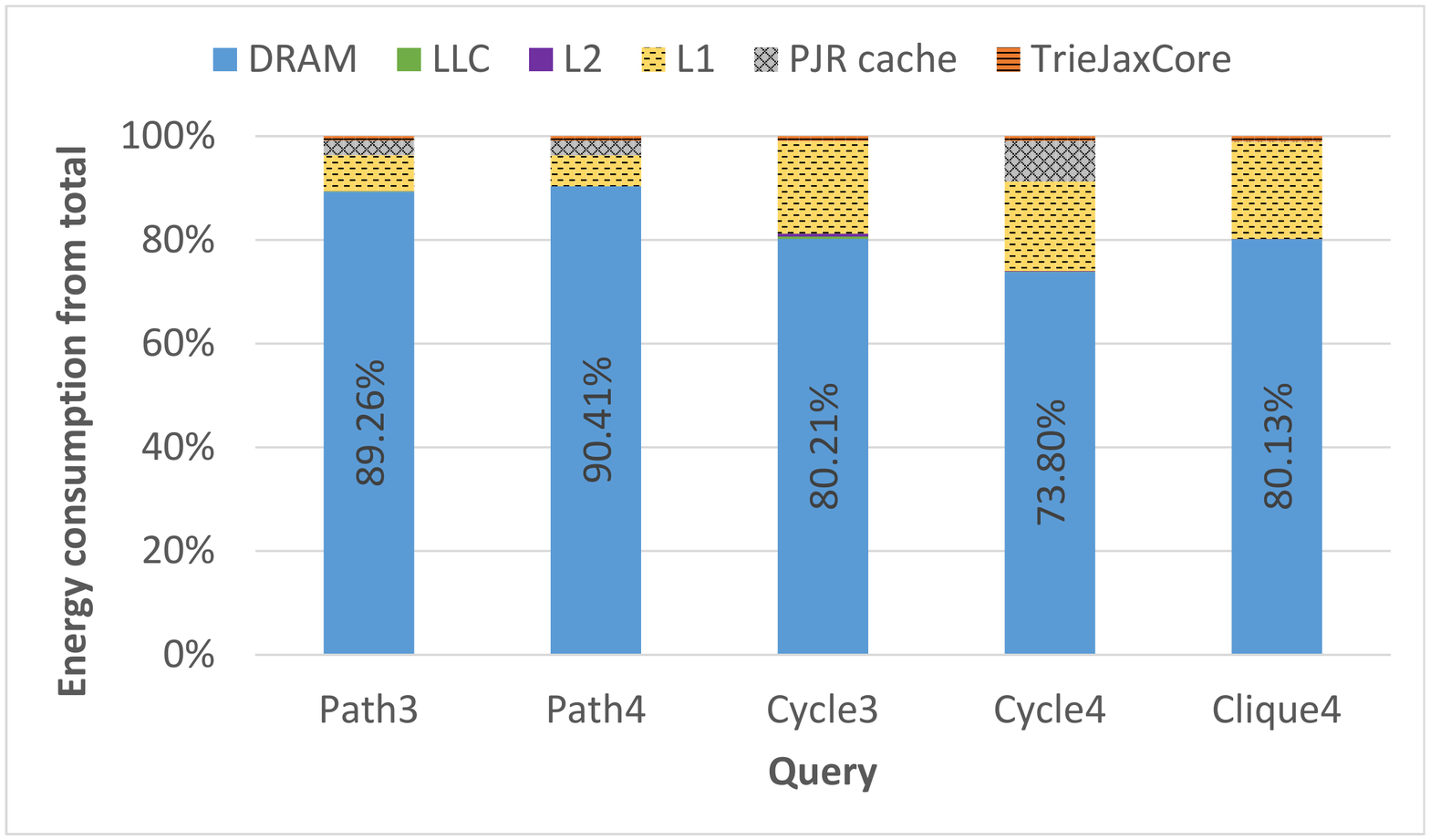}
    \caption{Average energy consumption distribution of \sysname for each query. Note that the energy consumption is completely dominated by the memory system. \label{fig:energyDist}}
\end{figure}

\begin{figure*}[t]
    \includegraphics[width=\textwidth, clip=true]{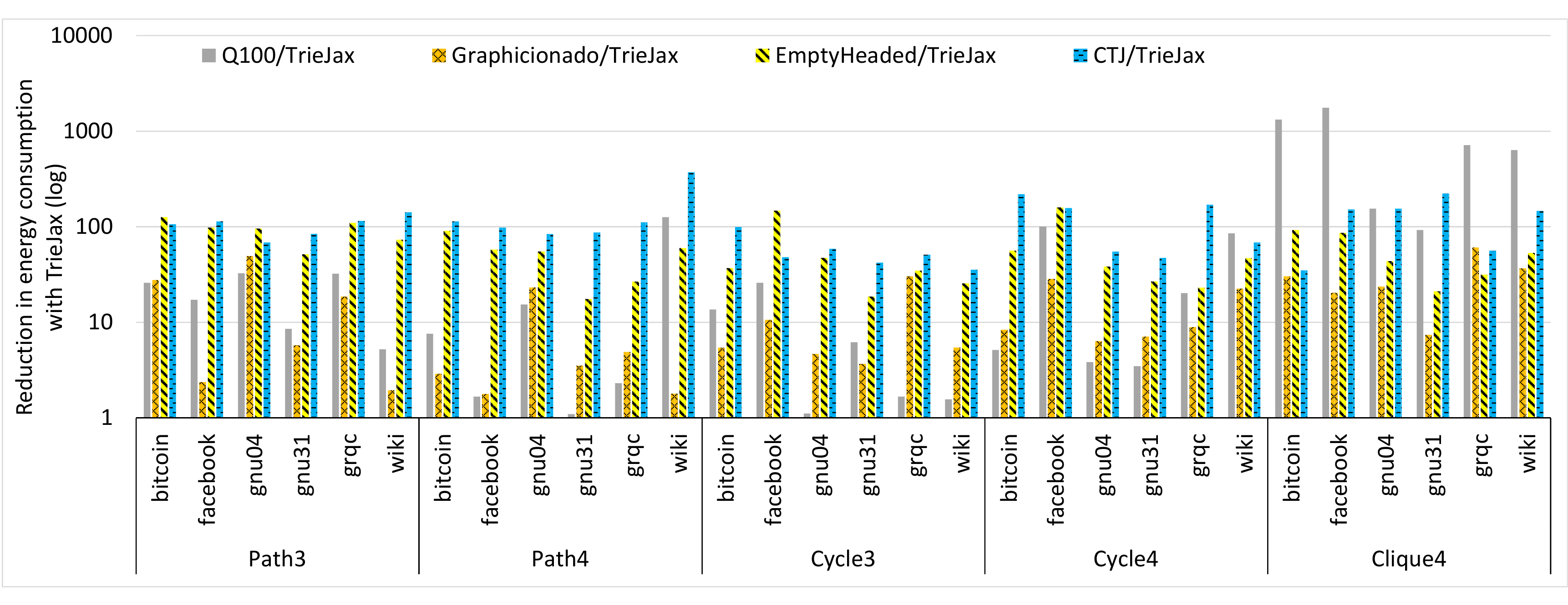}
    \caption{Reduction in energy consumption obtained with \sysname compared to the baselines. \label{fig:energy}}
\end{figure*}

\fig{fig:energyDist} presents the energy distribution in \sysname for the examined queries, averaged over the different datasets. 
The figure clearly shows that the lion's share of the energy consumed by \sysname goes to the memory system and only a fraction of the energy is consumed by the \sysname core logic.
The dominant memory component is DRAM, which consumes $74-90\%$ of the overall energy.
DRAM energy consumption is attributed to explicit read/write operations (mostly writing the final results to memory) and to implicit DRAM idle power (predominantly DRAM refresh operations).
The second dominant energy consumer is the L1 cache. Although the cache is read-only and is not involved in writing the final results to memory, it serves the trie traversals triggered by misses in the partial join results cache.
Finally, the \tirs is responsible for up to 7.8\% of the total energy consumption for the \e{cycle4} query. Note, however, that for \e{Cycle3} and \e{Clique4} queries there are no valid intermediate result caches and it does no use any energy.

We now turn to compare the energy consumption of \sysname versus that of the baseline systems.
\fig{fig:energy} presents the comparison.
Importantly, understanding that DRAM is by far the main energy consumer in \sysname enables us to explain its reduced energy consumption compared to the software and hardware baselines.

When comparing \sysname to the software baselines, it is not surprising to find that \sysname much more energy efficient.
\sysname is $110 \times$ more energy efficient on average than CTJ and $59 \times$ more efficient than EmptyHeaded.
This is attributed to three factors. 
The first factor contributing the \sysname's energy efficiency is the reduced energy consumption of its core logic, which is specifically designed to execute join operations and is thus much more energy efficient than a general-purpose core. This factor is slightly less dominant in the case of EmptyHeaded (which is almost $2 \times$ more efficient than CTJ), since EmptyHeaded uses efficient CPU SIMD operations.
The second contributing factor is the efficient \sysname's on-die caching of intermediate results and reduced  thrashing of the memory system caches, which dramatically reduces the energy consumption of its memory system.
The final contributing factor is the speedups obtained by \sysname, which dramatically reduce the DRAM's idle energy consumption. 

\sysname is also much more energy efficient than other hardware accelerators.
Specifically, \sysname consumes, on average, $179 \times$ and $15 \times$ less energy than Q100 and Graphicionado, respectively.
This improved energy efficiency is attributed to two factors. 
First, the use of a WCOJ algorithm that reduces the number of intermediate results and eliminates most of the expensive DRAM accesses executed by Q100 and Graphicionado.
Second, the faster run time of \sysname dramatically reduces the idle energy consumption of the DRAM, which is mostly the result of periodic DRAM refresh operations.

To summarize, we see that the energy efficiency of graph and database hardware accelerators is bounded by their memory system. Primarily, the number of explicit read/write operations affects both energy consumption and performance. But the performance impact of the memory system also increases DRAM's idle energy consumption.
\sysname's caching of internal results thus minimizes its reliance on the memory system, thereby reducing both its active memory energy  (read/write) and idle DRAM energy (implicit refresh).

\section{Conclusions \label{sec:conclusions}}
We presented \sysname, an on-chip domain-specific accelerator for graph operations specializing in graph pattern matching. 
It is driven by new advances in the database community and a plethora of new join algorithms that outperform traditional approaches. 
\sysname leverages the inherent concurrency of the algorithm for parallel execution and latency hiding of irregular memory accesses.
Furthermore, it integrates a specialized store for intermediate results that can drastically reduce recurrent computations.

We showed that \sysname outperforms state-of-the-art graph analytics and database accelerators by $7-63 \times$ on average, while consuming $15-179 \times$ less energy. 
\sysname further outperforms relational database management systems that use the modern join algorithms by $9-20 \times$ and consumes $59-110 \times$ less energy.

We plan to extend our accelerator to other important graph operations such as aggregations (e.g., triangle counting), and use novel algorithmic approaches to offer approximate estimations in a fraction of the time.





%
%
%

\bibliographystyle{IEEEtranS}
\bibliography{ref}

\appendix


\begin{figure*}
    \includegraphics[width=\textwidth, trim={0ex 10ex 0ex 10ex}, clip=true]{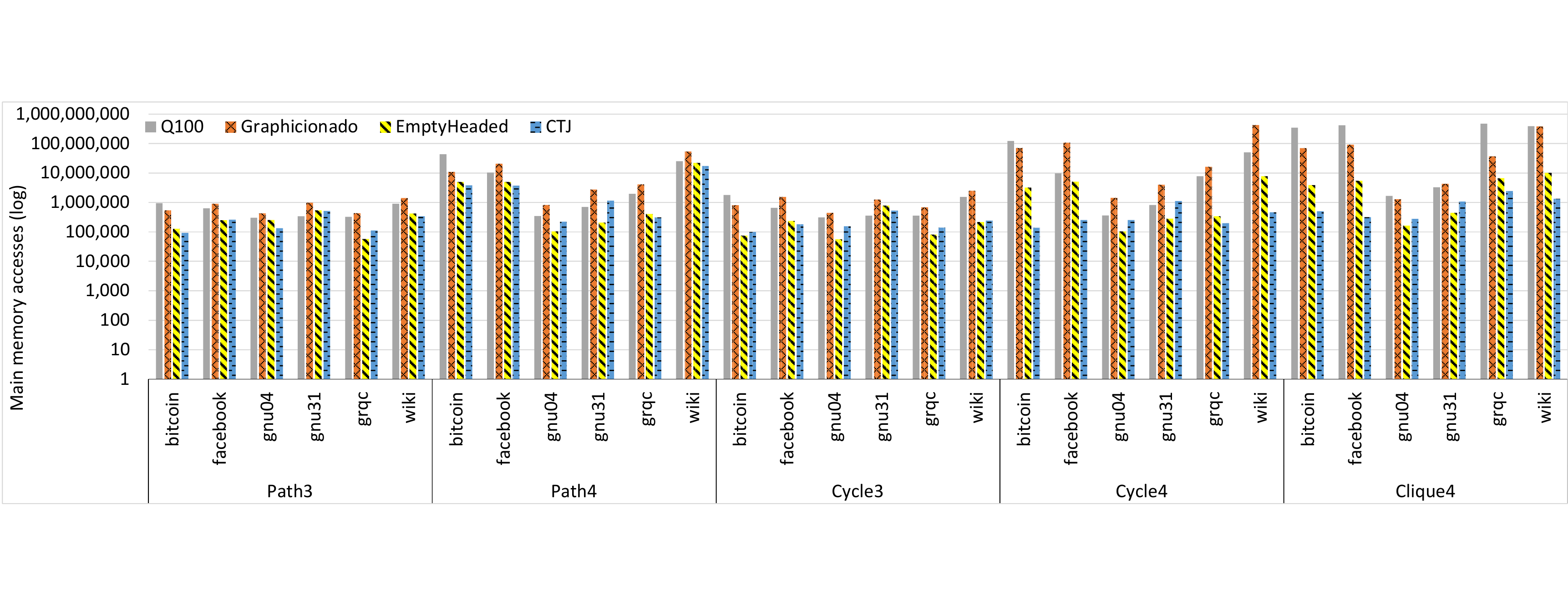}
    \caption{Number of memory accesses (log scale) for each baseline. \label{fig:memAccs}}
\end{figure*}
\begin{figure}
    \includegraphics[width=\columnwidth, trim={10ex 16ex 10ex 15ex}, clip=true]{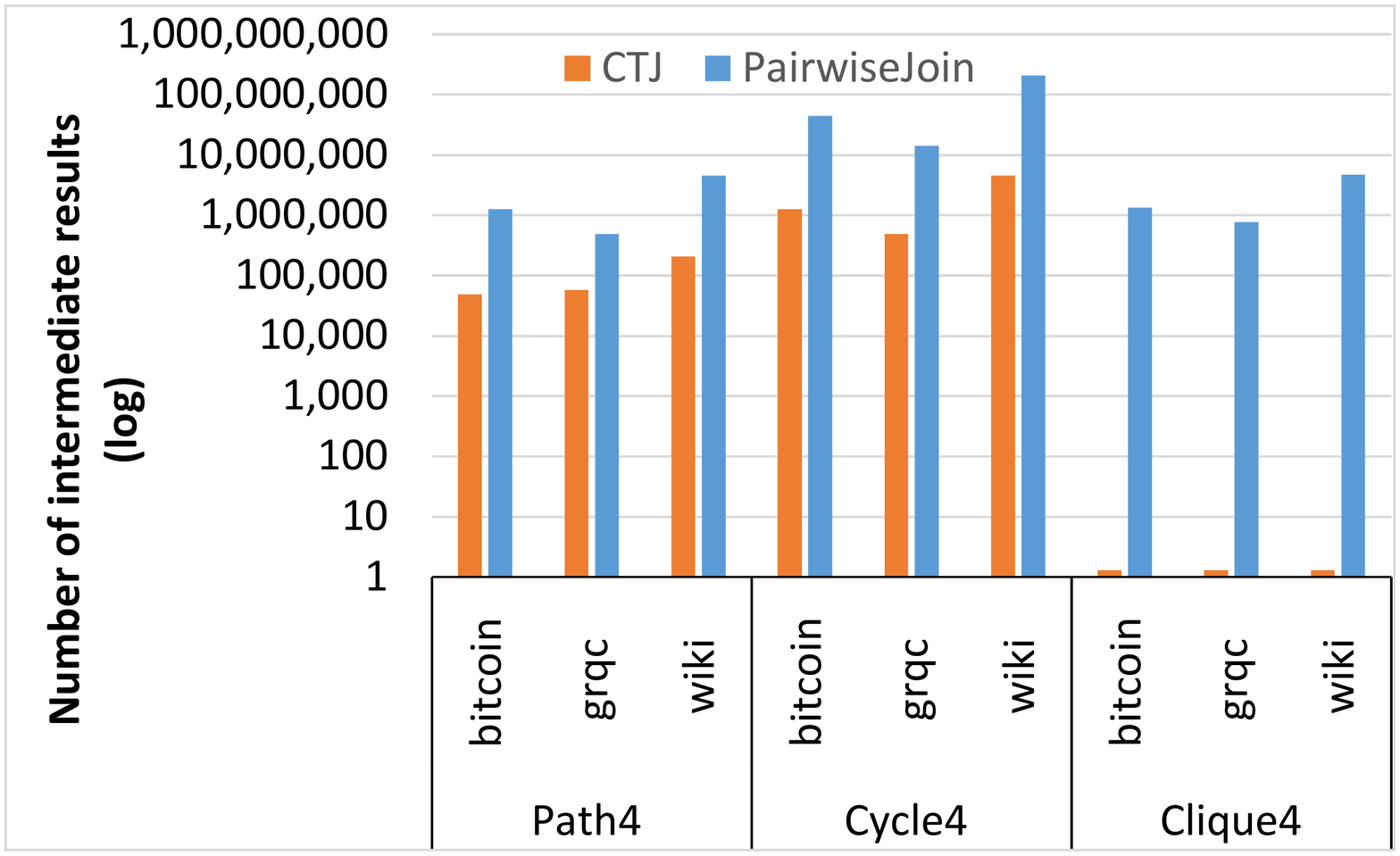}
    \caption{Number of intermediate results (log scale) being generated using CTJ compared to the pairwise join algorithm used in Q100. \label{fig:interRes}}
    \vspace{2ex}
\end{figure}

\section{Number of intermediate results}
The design of our system is based on a WCOJ algorithm. A key benefit is the reduction of intermediate results being generated during the computation of graph pattern matching queries. We illustrate the benefit in Figure~\ref{fig:interRes}, which compares the number of intermediate results (log scale) on a representative set of benchmarks using CTJ (WCOJ algorithm) and a pairwise join algorithm. The pairwise join algorithm is used in both Q100 and Graphicionado. 

On Path4 and Cycle4 queries, CTJ generates on average $18 \times$ and $36 \times$ fewer intermediate results than the pairwise join algorithm. While the pairwise join must generate all the intermediate results, CTJ only generates intermediate results that are part of the final result in compact caching structure.

Furthermore, CTJ only generates intermediate results that can later be reused to speedup the join. For example, CTJ does not generate any intermediate results for the Clique4 queries, because they cannot be reused.

\section{Main memory accesses}

Figure~\ref{fig:memAccs} shows the number of main memory accesses for each baseline on a logarithmic scale. The figure clearly shows that the WCOJ algorithms, namely EmptyHeaded and CTJ, generate fewer accesses to main memory than the traditional approaches, namely Q100 and Graphicionado. On average, CTJ generates $2.8 \times$ fewer memory accesses than EmptyHeaded, $47 \times$ fewer than Graphicionado, and $105 \times$ fewer than Q100. This matches our motivation for using a WCOJ-based solution for our system.

\end{document}